\documentclass[letterpaper,twocolumn,10pt]{article}
\usepackage{usenix}

\usepackage{amsmath}
\usepackage[ruled,linesnumbered]{algorithm2e}
\usepackage{amsfonts}
\usepackage{xspace}
\usepackage{xcolor}
\usepackage{graphicx}
\usepackage{hyperref}
\usepackage{subcaption}
\usepackage{multirow,multicol}
\usepackage{booktabs}
\usepackage{xurl}

\newcommand{\IR}{\mathbb{R}}
\newcommand{\G}{\mathcal{G}}
\newcommand{\V}{\mathcal{V}}
\newcommand{\E}{\mathcal{E}}

\usepackage{mathtools}
\DeclarePairedDelimiter\ceil{\lceil}{\rceil}

\newcommand{\mat}[1]{\begin{bmatrix}#1\end{bmatrix}}

\definecolor{mulberry}{rgb}{0.77, 0.29, 0.55}

\newcommand{\linebreakand}{%
  \end{@IEEEauthorhalign}
  \hfill\mbox{}\par
  \mbox{}\hfill\begin{@IEEEauthorhalign}
}

\newcounter{ObservationCounter}

\newcommand{\name}{{\textsc{CyberGFM}}\xspace}
\newcommand{\comment}[1]{}

\newif\ifsubmit
\submitfalse 
\ifsubmit
    \newcommand{\isaiah}[1]{}
    \newcommand{\bernardo}[1]{}
    \newcommand{\ramiro}[1]{}
    \newcommand{\ben}[1]{}
    \newcommand{\howie}[1]{}
    \newcommand{\pradeep}[1]{}
    \newcommand{\outline}[2]{}

    \newcommand{\rev}[2]{}
\else
    \newcommand{\isaiah}[1]{\textcolor{mulberry}{\textbf{Isaiah: #1}}}
    \newcommand{\bernardo}[1]{\textcolor{teal}{\textbf{Bernardo:} #1}}
    \newcommand{\ramiro}[1]{\textcolor{blue}{\textbf{Ramiro: #1}}}
    \newcommand{\ben}[1]{\textcolor{purple}{\textbf{Ben: #1}}}
    \newcommand{\howie}[1]{\textcolor{red}{\textbf{Howie: #1}}}
    \newcommand{\pradeep}[1]{\textcolor{green}{\textbf{Pradeep: #1}}}
    \newcommand{\outline}[2]{\textcolor{blue}{\##1:~}\textcolor{cyan}{#2}}

    \newcommand{\rev}[2]{\textcolor{green}{Review #1: ``#2"}}
\fi

\begin{document}

\title{\Large \bf CyberGFM: Graph Foundation Models for \\ Lateral Movement Detection in Enterprise Networks}

\author{
    {\rm Isaiah J. King~\footnotemark[1]~~\footnotemark[2]}
    \and 
    {\rm Bernardo Trindade~\footnotemark[2]}
    \and 
    {\rm Benjamin Bowman~\footnotemark[1]}
    \and 
    {\rm H. Howie Huang~\footnotemark[1]~~\footnotemark[2]}
    \and \\
    {\rm \footnotemark[1]~~Cybermonic LLC. McLean, VA, USA.} \\
    {\rm \footnotemark[2]~~The George Washington University. Washington, DC, USA.}
}

\maketitle

\begin{abstract}
Representing networks as a graph and training a link prediction model using benign connections is an effective method of anomaly-based intrusion detection.
Existing works using this technique have shown great success using temporal graph neural networks and skip-gram-based approaches on random walks. 
However, random walk-based approaches are unable to incorporate rich edge data, while the GNN-based approaches require large amounts of memory to train. 
In this work, we propose extending the original insight from random walk-based skip-grams--that random walks through a graph are analogous to sentences in a corpus--to the more modern transformer-based foundation models. 
Using language models that take advantage of GPU optimizations, we can quickly train a graph foundation model to predict missing tokens in random walks through a network of computers. 
The graph foundation model is then finetuned for link prediction and used as a network anomaly detector. 
This new approach allows us to combine the efficiency of random walk-based methods and the rich semantic representation of deep learning methods.
This system, which we call {\name}, achieved state-of-the-art results on three widely used network anomaly detection datasets, delivering a up to 2$\times$ improvement in average precision.
We found that {\name} outperforms all prior works in unsupervised link prediction for network anomaly detection, using the same number of parameters, and with equal or better efficiency than the previous best approaches. 
\end{abstract}

\section{Introduction}
\label{sec:intro}
Lateral movement is a critical stage of early compromise. 
In this phase, the attackers have entered the network, and try to pivot through hosts to continually gain privileges, before reaching their target. 
Under the Mitre ATT\&CK framework~\cite{mitre_attack}, lateral movement is the final stage before the attackers can achieve their objectives: data collection, command and control (C2), or impact. 
Clearly, detecting it early is crucial for the safety of any network. 

Prior works in lateral movement detection achieve this by representing the network as a graph. 
It has been shown that lateral movement in a network graph manifests as low-probability edges, which can be detected using graph link prediction~\cite{bowman2020,pikachu,euler,argus}.
These approaches  train on a benign period of network data, and use this to establish a baseline of normal activity. 
During inference, new edges with low probability are classified as anomalous and indicative of lateral movement. 
While the prior works have made impressive improvements in precision, generating models with very low false positive rates, the massive size of network data necessitates heavy computation. 
Random walk-based works like~\cite{bowman2020,pikachu} are less affected by this issue if efficient sampling techniques are used, but only use shallow embedding networks, leading to worse performance. 
On the other hand, deep learning provides greater expressiveness in node representation, meaning greater precision and recall, but doing so on such large graphs is challenging. 
Prior works~\cite{euler,argus} account for this by distributing graph neural networks (GNNs) across multiple CPUs, but this comes at the cost of communication overhead and cannot take advantage of the acceleration offered by GPUs.

The key insight behind this work is to take the memory efficiency of random walk approaches, and analyze them using more powerful models. 
The DeepWalk paper~\cite{deepwalk} showed that random walks through graphs are analogous to sentences. 
With this assumption in mind, they analyzed them with word2vec~\cite{w2v}, the best available natural language processing (NLP) model at the time. 
Since 2013, modern NLP models have moved past the context-free embeddings of word2vec, and instead analyze each word in the context of full sentences.
Large language models (LLMs) like GPT~\cite{gpt} and BERT~\cite{bert} are instead built on multiheaded self-attention networks~\cite{attn}, mechanisms that look at the full sentence, and embed each word in the context of every other word present. 
In this work, we try to answer the question: How can graph analysis take advantage of recent advances in NLP?

To this end, we propose using random walks through a graph of network activity to train a large language model. 
Like the prior random walk-based methods for lateral movement detection~\cite{bowman2020, pikachu, rabbani2024}, paths through the network are treated as sentences, and used to train an NLP model. 
The crucial difference is that the NLP model we select is much more powerful than shallow skip-gram networks that power word2vec. 
We train both a BERT and a GPT model using benign network activity as its corpus, allowing the model to learn the likelihood of any given link, given the context of a full walk through the neighborhood.
This allows the model to capture distances between nodes, as well as the directionality of paths between nodes, in a way that earlier language models are unable to encode.
Using scheduled masked token prediction on random walks, we train a model to predict missing nodes in random walks. 
The result of this unsupervised pretraining is a graph foundation model (GFM) with a generalized understanding of common paths through the underlying graph. 
Importantly, the transformers that power these models are designed for GPU acceleration.
This allows our models to train and perform inference with high efficiency without the need for multiprocessing. 

As we will show, the GFM can be used as an anomaly detector on its own. 
Given the source and destination nodes $u,v$ for an edge we wish to test, an (optional) edge feature $\mathbf{x}_{(u,v)}$, and a random walk terminating $\{n_1,x_{e_1},..., u,x_{e_{(u,v)}},\texttt{[MASK]}\}$ the GFM will output the likelihood that $v$ is the masked token in the sequence. 
However, the GFM can be further \textit{fine-tuned} for our specific objective. 
Passing the GFM through an additional round of training to classify sequences as malicious or benign further improves its precision. 
Using this method, we train a highly precise link prediction model with comparable parameters to existing works that achieves a more than 2x improvement in average precision. 
This new approach allows us to combine the efficiency of random walk-based methods and the rich semantic representation of deep learning methods. 
This model, which we call {\name}\footnote{
Source code available at \url{https://github.com/cybermonic/CyberGFM}
}, outperforms state-of-the-art unsupervised lateral movement detection models by a wide margin. 

The primary contributions of this paper are as follows: 

\begin{itemize}
    \item \textbf{A novel system of lateral movement detection}: 
    To the best of our knowledge, this is the first approach to model lateral movement detection as next token prediction with an LLM architecture. 
    The LLM architecture has proven itself to be highly effective in the NLP space, and our approach extends it to both graphs and cybersecurity analysis. 
    \item \textbf{State-of-the-art anomaly detection}:  
    We evaluate our model against four recent anomaly-based intrusion detection systems on three cybersecurity datasets. 
    The fine-tuned GFM outperforms all prior works on all three datasets, demonstrating the power of this approach. In particular, our model achieves a 0.76 AP score on the LANL dataset, which is more than double the previous best score. 
    \item \textbf{GFMs for cybersecurity applications}:
    While prior works on GFMs rely on expensive subgraph sampling methods, ours adopts the DeepWalk approach of assuming random walks are analogous to sentences. This allows for more efficient pretraining and the ability to process much larger graphs. Additionally, we show that models pretrained with our approach can be finetuned, and continue to make large performance improvements, while models trained directly on the fine-tuning task are slower to converge, and frequently collapse. 
\end{itemize}   
\begin{figure*}
    \centering
    \begin{subfigure}{0.48\linewidth}
        \centering
        \includegraphics[width=\linewidth]{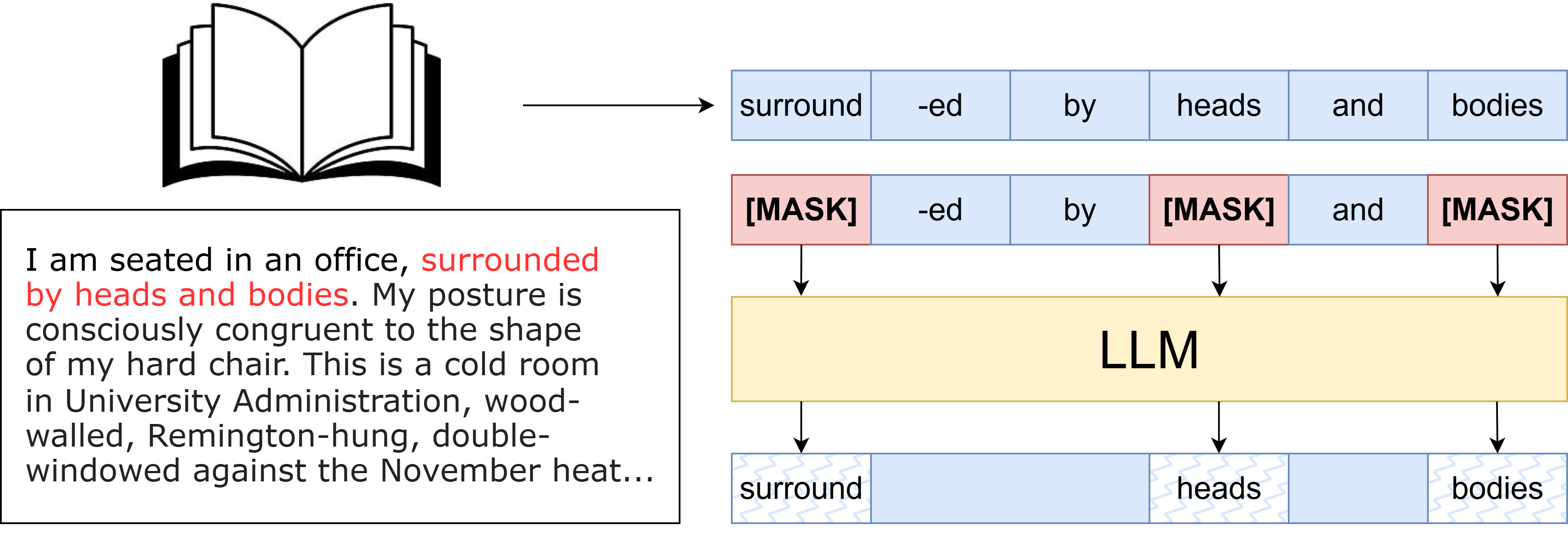}%
        \subcaption{Traditional LLM pretraining}
        \label{subfig:llm}
    \end{subfigure}%
    \hspace{0.04\linewidth}%
    \begin{subfigure}{0.48\linewidth}
        \centering
        \includegraphics[width=\linewidth]{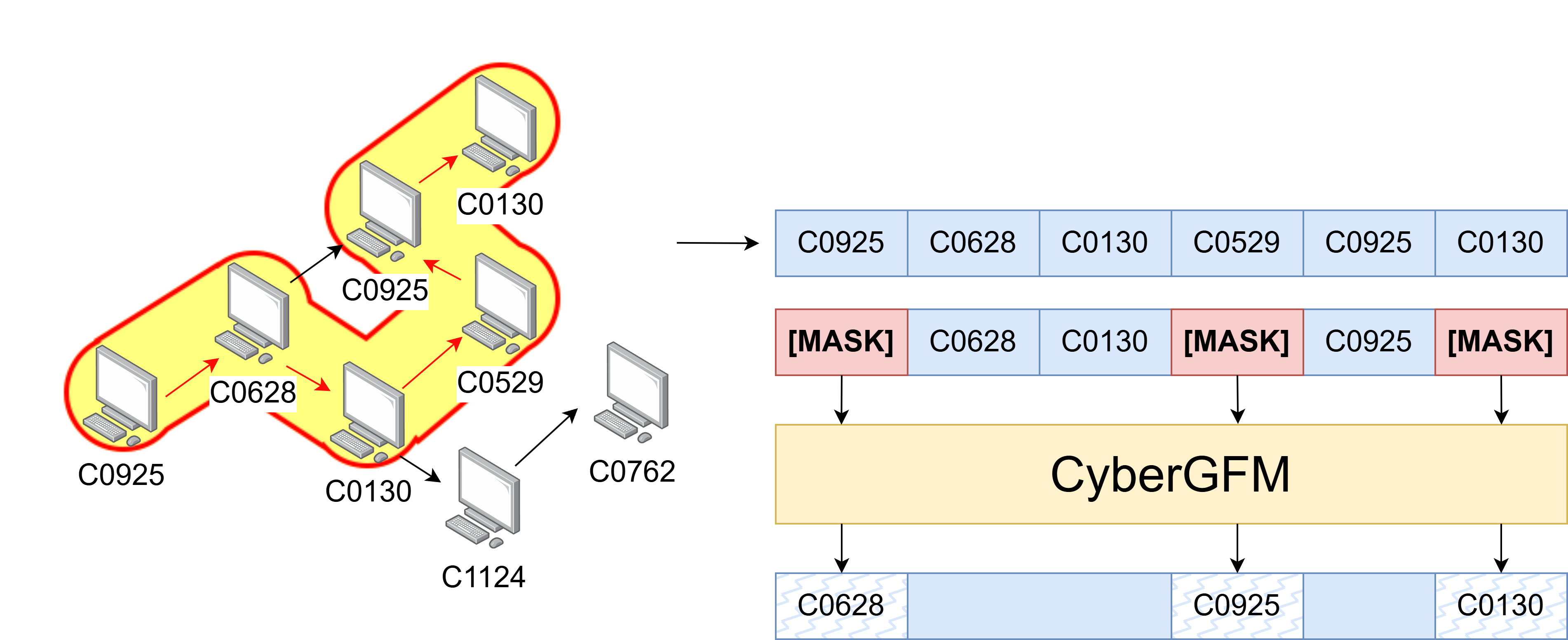}
        \subcaption{Proposed GFM pretraining}
        \label{subfig:gfm}
    \end{subfigure}
    \caption{(a) The traditional process to pretrain LLMs samples sentences from a large corpus, masks out tokens, then optimizes the model based on its ability to predict masked tokens in the output sequence. (b) Our proposed method for GFM pretraining follows this same paradigm, using random walks through the graph as ``sentences" where tokens represent node IDs and edge features.}
    \label{fig:llm_vs_gfm}
\end{figure*}

\section{Background}
\label{sec:bg}

\subsection{Network Log Data as a Graph}
A graph $\G = \{\V, \E\}$ is defined as a set of entities, or nodes $\V$, and a set of relationships between them, or edges $\E \subseteq \{(u,v) \mid u,v \in \V\}$. 
Additionally, edges may have features $\mathbf{x}_{(u,v)}$ containing important information about the interaction that occurred between the entities.
Importantly, one edge feature that all graphs we consider will have is time: the instant that the edge between $u$ and $v$ occurred. 

The activity of a computer network lends itself naturally to being expressed as a graph. 
Hosts, users, files, etc., may be represented as the set of nodes, with important features such as their type as their features. 
Interactions between these entities, such as communication, authentication, file creation, etc., are naturally represented as a directed edge between them. 
In this work, we will represent logs of network flows and authentications between computers as graphs with attributed edges. 
The edges all have a feature for time, and when available, features about the kind of communication that occurred between the entities, as well as any other available information, such as the number of packets sent, flow duration, and number of bytes sent. 

\subsection{Link Prediction for Lateral Movement Detection}
Lateral movement is the phase of an attack where the adversary pivots from their initial foothold to try to reach their target. 
This phase of the attack involves exploring and compromising new hosts across a network. 
When networks are represented as graphs, lateral movement manifests as anomalous edges~\cite{bowman2020}. 
This is because the path from a vulnerable entry point to whatever critical asset the attacker desires will involve connections that are unlikely to appear during normal operations (e.g., a user in the HR subnet authenticating as an admin on the domain controller). 

Link prediction is the process of finding a function $f(u,v)$ that determines the likelihood of an edge between nodes $u$ and $v$. 
As prior work~\cite{argus} identified, edge features such as protocol, number of bytes sent, etc., can provide valuable data about how likely a particular edge is to occur. 
Thus, we adjust the function to be $f(u,v \mid \mathbf{x}_{(u,v)})$ where $x_{(u,v)}$ is a vector of features associated with observed edge $(u,v)$. 
This allows our model to distinguish between, e.g., a user requesting a certificate for themselves from the Active Directory Certificate Service, vs a user requesting a certificate for the domain admin during an ESC1 attack~\cite{esc1}. 
For these reasons, we formulate the task of lateral movement detection as a link prediction problem. 
We aim to detect both anomalous connections and the anomalous \textit{context} of those connections. 

\subsection{Foundation Models}
A foundation model (FM) is one that first undergoes unsupervised training to learn a general task, then is applied to a different, more specific task in the same domain~\cite{fm_survey}.
Often, FMs undergo an additional round of supervised training called \textit{fine-tuning} before they are applied to their narrower task. 
The most familiar form of FMs are large language models, though FMs exist for image~\cite{image_fm}, video~\cite{video_fm}, audio~\cite{audio_fm}, and graph~\cite{graphgpt} modalities, as well as many more~\cite{fm_survey2}. 
The pretraining task for large language models is to predict the next or a missing token in an input sequence. 
The resulting FM can then be fine-tuned for sentiment analysis, classification, or any other number of tasks. 

In NLP, a transformer encoder~\cite{attn} is used as the main architecture. 
They are pretrained for next token prediction (NTP), or (scheduled) masked token prediction (SMTP) in a sequence~\cite{bert}. 
In this work, we will focus on FMs trained via SMTP. 
Figure~\ref{subfig:llm} illustrates this process in an LLM setting. 
Sentences are sampled from a large corpus of text, turned into a sequence of tokens that map to words or clusters of characters, and then some percentage of the sequence is masked out. 
This sequence is passed through a transformer encoder, which outputs the likelihood that input at position $i$ was a given token. 
This seemingly simple task when given enough data can train a model general enough that it can easily transfer its knowledge to other NLP tasks. 
The model, after the unsupervised pretraining session, can now be optimized for new problems and achieve much better scores than models that have not undergone this pretraining step~\cite{why_does_pretraining_work,pretrain_v_from_scratch}. 
\section{System Design}
\label{sec:method}

\begin{figure*}
    \centering
    \begin{subfigure}{0.4\linewidth}
        \centering
        \vspace{1em}
        \includegraphics[width=0.9\linewidth]{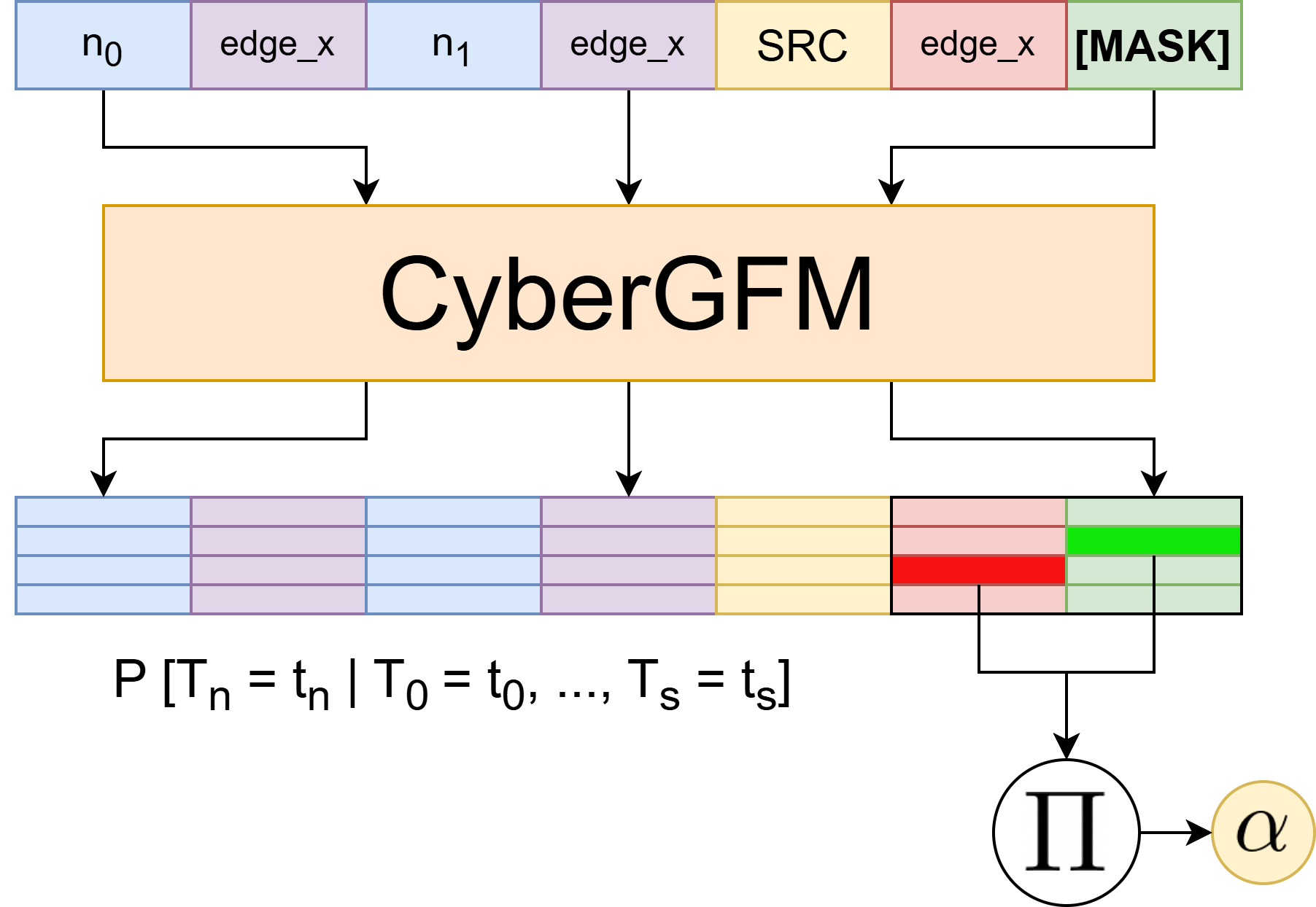}
        \subcaption{Link prediction-based fine-tuning}
        \label{subfig:lp_based}
    \end{subfigure}
    \hfill
    \begin{subfigure}{0.48\linewidth}
        \centering
        \includegraphics[width=0.9\linewidth]{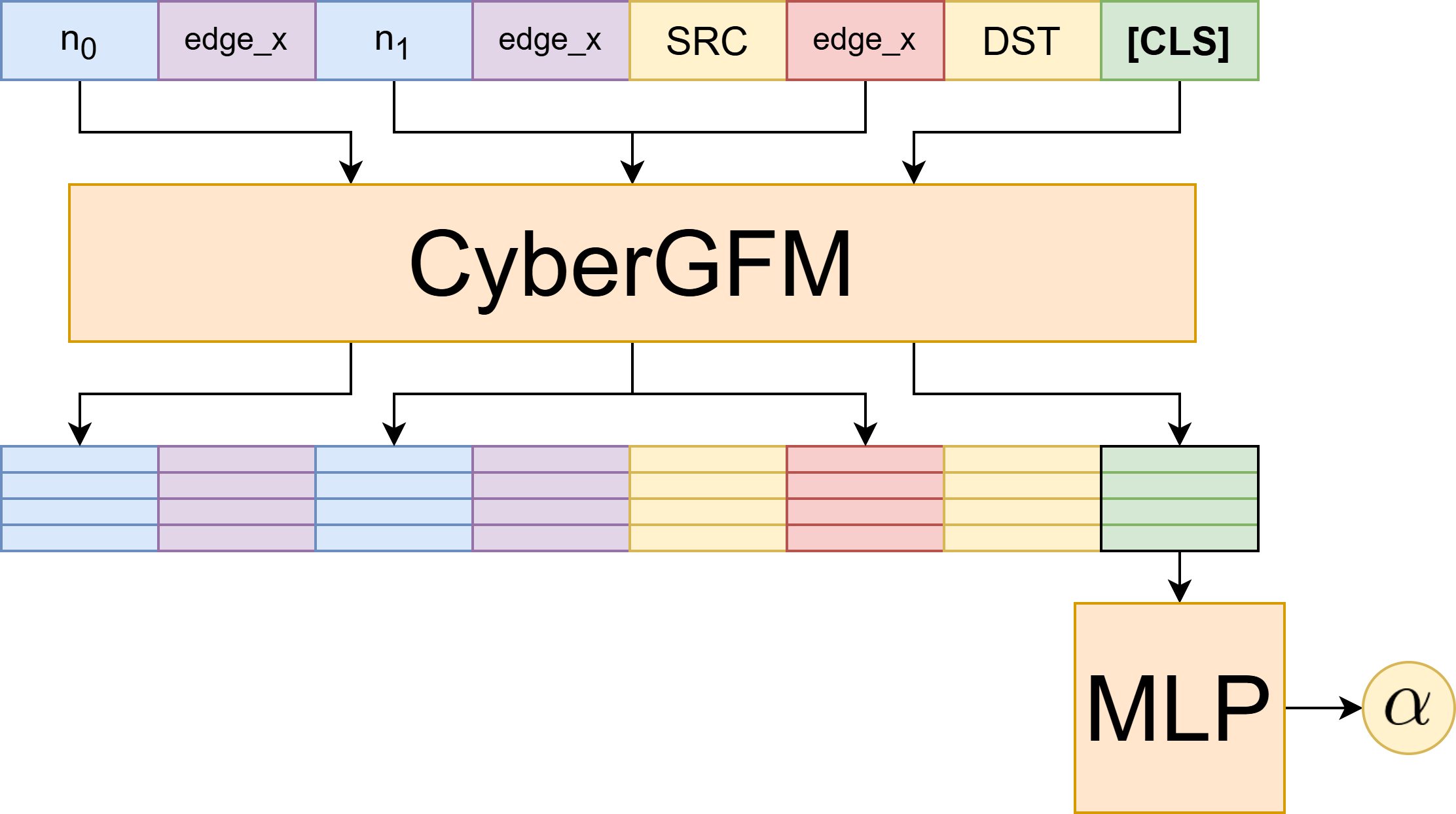}
        \subcaption{Classification-based fine-tuning}
        \label{subfig:cls_based}
    \end{subfigure}
    \caption{Fine-tuning options}
    \label{fig:finetuning}
\end{figure*}

Our system, {\name}, trains a foundation model for graphs of network security logs. 
The system can be broken into three steps, which this section will cover in detail: graph construction, pretraining, and fine-tuning. 
Before we describe the steps, we provide a brief overview of our method, and introduce the concept of graph foundation models, extending traditional FMs. 

\subsection{Graph Foundation Models}
Foundation models have shown success in their application to a wide variety of tasks in NLP. 
This has led researchers to ask whether the pretraining-fine-tuning pipeline will work in other domains, such as images, robotics, and formal logical reasoning~\cite {fm_survey,fm_survey2}. 
When applied to the domain of graph-structured data, we refer to these models as \textit{graph foundation models} (GFMs). 
Unlike the ubiquity of FMs in the NLP space, GFMs are relatively new. 
As such, there still is no agreed-upon best method of pretraining them, or even a preferred architecture.
Approaches range from pretraining large GNNs on metagraphs of relations~\cite{ultra}, to the more familiar next/masked token prediction on sampled subgraphs~\cite{graphbert,graphgpt}. 

In this work, we propose an extension of the paradigm set forth by DeepWalk~\cite{deepwalk}: random walks through a graph, in addition to being far more efficient to generate than subgraph samples, are analogous to sentences in a corpus. 
Thus, random walks through a network of computers are used as the input to train {\name}.
To our knowledge, this is the first work to train a GFM in this manner and the first application of GFMs for anomalous lateral movement detection.

During pretraining (Subfigure~\ref{subfig:gfm}), we first generate random walks through a graph of benign activity. 
As illustrated in Figure~\ref{fig:llm_vs_gfm}, this is a natural extension of FMs in the NLP domain. 
The training pipeline is nearly identical; the only difference is that the input tokens represent random walks through a network rather than groupings of characters in a corpus.
These random walks are then turned into a series of tokens, with some proportion of the tokens masked out or perturbed before the walks are fed into {\name}. 
The model is then optimized to predict the most likely tokens that were masked out, given the context of the complete random walk. 

In our model, a token is any discrete entity one wishes to encode.
In NLP, tokens map to sets of characters or words~\cite{tokenizer}; in this work, tokens map to nodes in a graph and edge features.  
Given $\mathcal{T}$ possible values for tokens, let $\mathbf{T}$ denote a $|\mathcal{T}| \times d$ trainable parameter. 
A sequence-to-sequence encoder will take a $B\times S$ sequence of tokens as input. 
The tokens are used as row indexes for $\mathbf{T}$, such that a sequence of tokens $\{t_0, ..., t_n\}$ will be represented as a matrix built from the corresponding rows in $\mathbf{T}$. 
This extracted matrix is then combined with a function to encode token order before it is passed through a series of transformers. 
The output of the final transformer is passed through a fully connected layer to produce a $B \times S \times |\mathcal{T}|$ tensor, representing the probability that the input token at position $S$ is token $\mathbf{T}_i$. 

In traditional NLP, this process has been shown to produce highly generalizable models that can be fine-tuned to solve other tasks over the same space of input sequences~\cite{why_does_pretraining_work}. 
In this work, we extend this notion to lateral movement detection. 
We posit that by pretraining a GFM on benign sequences of connections in network logs, the model learns a distribution of normal authentications. 
On its own, the foundation model can be used to predict the likelihood of a given connection occurring by providing $\mathcal{S} = [\texttt{[nid:SRC]}, \texttt{[MASK]}]$ as input, and using $P(\mathcal{S}[1] = dst \mid f(\mathcal{S}))$, where $f(\cdot)$ is the pretrained GFM. 
By continuing to fine-tune the model on pairs of known benign edges, we will show that the final model can classify network activity as malicious or benign with high precision. 

More specifically, we evaluate two methods of unsupervised fine-tuning: link prediction-based (Subfigure~\ref{subfig:lp_based}) and classification-based (Subfigure~\ref{subfig:cls_based}).
Both methods are trained to classify a single edge $(u,v)$ as anomalous or benign, given a random walk terminating at $..., u$ or $..., u, \mathbf{x}_{(u,v)}$ if the graph has edge features. 

In link prediction-based fine-tuning, a \texttt{[MASK]} token is appended to the end of each walk. 
The product of the likelihoods of $\mathbf{x}_{(u,v)}$ and $v$ for the last positions in the sequence is used as the anomaly score for the edge. 
On the other hand, in classification-based fine-tuning, a special, classification token, \texttt{[CLS]}, is appended to the end of the sequence. 
The output of the sequence at the position of this token is passed through a fully-connected neural net, and that network's output is used as the anomaly score for the edge. 

\subsection{Graph Construction}
Before we dive into GFM pretraining, we will discuss briefly how we use graphs to model the network logs. 
The graph construction phase varies between datasets, but in general, it involves finding some mapping that extracts two entities from each line of log data and stores them as an edge in a graph. 

Generally speaking, we represent networks as temporal graphs where nodes represent hosts, and edges represent a communication event occurring between them. 
Each edge is attributed with a timestamp denoting when it occurred.
Mathematically, the graph can be represented as the set of nodes $\V$ and attributed edges $\E$, where $\mathcal{E} \subset \{(u,v,t) \mid u,v \in \V, t\in \IR^+\}$. 
We represent graphs in compressed sparse row (CSR) format to quickly find each node's neighbors during random walks and to save disk space. 

Graphs may optionally have edge features in addition to their timestamps $f: \E \rightarrow \IR^d$. 
These features may represent specific details about the communication occurring between the two entities the graph represents. 
Details of the specific edge features used for each dataset are provided in Subsection~\ref{subsec:datasets}.

\subsection{Pretraining}
Following the intuition of the original DeepWalk paper~\cite{deepwalk}, if random walks through a graph can be treated as sentences, then they can be analyzed with NLP approaches. 
Our approach involves generating (temporally biased) random walks through the graph representing the network, and using these walks as the ``corpus" by which we train an LLM. 
The full pretraining procedure consists of (1) sampling (temporal) random walks from the graph, (2) masking or replacing random nodes from these sequences, and finally (3) using BERT to predict the missing tokens. 

\textbf{Random Walks} are generated in parallel for each node in a given batch. 
Because the graph is stored in compressed sparse row (CSR) format, it is efficient to find the neighbors of each node. 
The graph is constructed such that neighbors are ordered by time within the CSR rows, so it is also efficient to temporally bias this search. 
We evaluate GFMs pretrained using standard random walks, as well as temporally biased random walks~\cite{ctdne}. 
To generate these walks, we modify the standard random walk algorithm as shown in Algorithm~\ref{alg:trw}. 

\begin{algorithm}[htbp]
\caption{Temporal Random Walk Generation}
\label{alg:trw}
\SetKwInput{KwInput}{Input}
\KwInput{Target node: $n$, CSR graph: \{\texttt{idxptr}, \texttt{col}, \texttt{ts}\}}
$i\gets 0$; $t_n \gets 0$; $e_n \gets 0$; walk $\gets$ []\;

\While{len(walk) $<$ walk length}{
    start $\gets$ \texttt{idxptr[$n$]}; end $\gets$ \texttt{idxptr[$n+1$]}\;
    \uIf{$e_n == -1$}{
        start = end;
    } 
    \Else {
    \tcc{First occurrence of edge with $t_e \ge t_n$}
        start = binary\_search(\texttt{ts}, start, end, $t_n$)\;
    }
    \BlankLine
    \uIf{start == end} {
    \tcc{Node has no neighbors}
        $e_n = -1$\;
    }\Else{
        $e_n$ = start + Random() * (end-start)\; 
        $n \gets\texttt{col[}e_n\texttt{]}$\;
    }
    walk.append($n$); $t_n \gets \texttt{ts[}e_n\texttt{]}$\;
}
\end{algorithm}

First, the range in the column vector containing neighbors of the input node $n$ is accessed (line 3). 
All values of \texttt{col[idxptr[n] : idxptr[n+1]]} are neighbors of node $n$, and all times in \texttt{ts[idxptr[n] : idxptr[n+1]]} are the times at which those edges occurred. 
Next, if the node has neighbors, we find the lowest edge index \texttt{i} between \texttt{start} and \texttt{end} where $\texttt{ts[i]} \ge t_n$. 
This ensures that each of the possible next neighbors to walk to has an edge from the source node later in time than the edge that brought us here (line 7).  
Note that skipping this line is sufficient to produce non-temporally biased random walks. 
Finally, given the range of allowed edges to take for its next step, the algorithm randomly selects one (lines 12--13) and updates the last edge's timestamp (line 15). 

\begin{figure}[t]
    \centering
    \includegraphics[width=\linewidth]{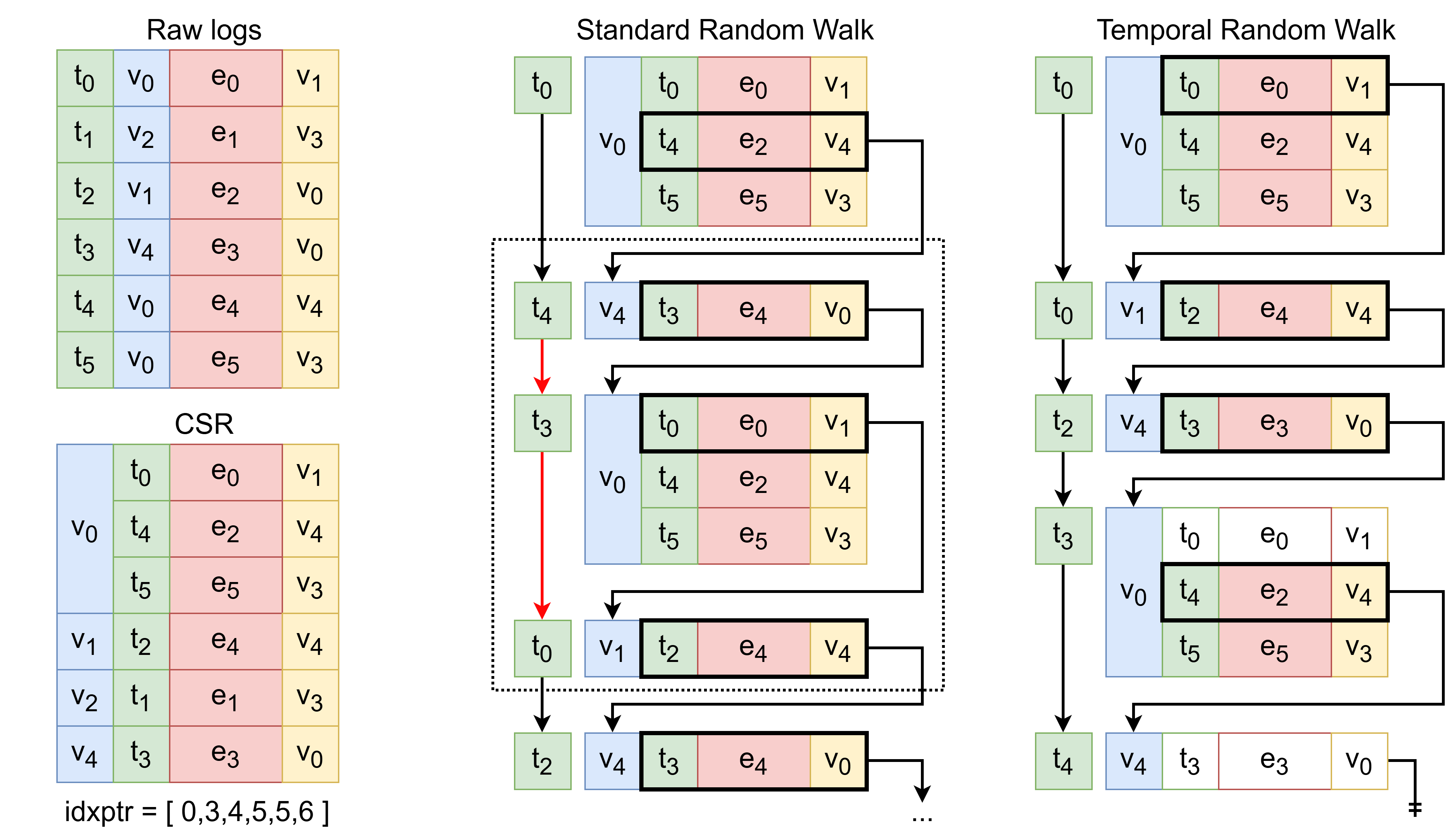}
    \caption{An example of going from raw log files to a CSR representation, then extracting random walks. For the standard random walk shown in the middle, the area in the dotted region ignores the temporal constraint. }
    \label{fig:rw_example}
\end{figure}

This approach is similar to the one proposed in prior work~\cite{ctdne} and applied by~\cite{pikachu}. 
Both works note the importance of only allowing random walks to occur forward in time; this way, causal relations between nodes are not lost by the compression into a static graph. 
To further illustrate this difference, Figure~\ref{fig:rw_example} shows two example random walks, a standard random walk, and a temporally biased one. 
In the standard random walk, at any given point, all neighbors of the source node can be selected as the next edge in the walk; in the temporally biased walk, only edges that occur later in time can be selected (edges that occurred in ``the past" are illustrated as blocks without color). 
In our experiments, we evaluate models trained with both temporally biased, and traditional random walks to more fully evaluate this assertion. 

\textbf{Tokenization}: Random walks are converted to sequences of tokens to be analyzed by the GFM. 
In this work, as graphs to be analyzed are smaller than 20k nodes, node IDs can be tokenized directly. 
For larger graphs, node IDs may be subdivided into multiple tokens, as was done by~\cite{graphgpt}. 
Additionally, for graphs with edge features, those features are also converted into discrete tokens and included in the random walk. 
In our experiments, categorical features are tokenized directly. 
For integer edge features, values 0-100 were tokenized directly, and larger values were tokenized as $\ceil{\log_{10}(x_{e_{i,j}})}$, where $x_{e_{i,j}}$ is the $j$th feature of edge $e_i$. 
Some datasets also included floating-point edge features, typically representing the average counts of events during a specific network flow. 
For values greater than 1, we cast them to integers and tokenize them as before. 
For values less than 1, we considered multiples of 0.01 between $[0.00, 0.99]$ inclusive and truncated the values beyond two decimal places. 

For example, a path $\{n_0, e_0, n_1, e_1, n_2\}$ where $e_0 = \mat{\text{`UDP'} & 0.5687 & 19}$, and $e_1 = \mat{\text{`TCP'} & 2.01 & 1,123}$ would be converted to the sequence of tokens: \texttt{[[nid:0], [ef:UDP], [ef:0.56], [ef:19], [nid:1], [ef:TCP], [ef:2], [ef:10\string^4], [nid:2]]}. 
We use the same tokens for all features. That is, if an edge has features $\begin{bmatrix}1 & 1\end{bmatrix}$, both instances of 1 are represented using the same token. 
We rely on the model to use the positional encodings of each token to infer that these symbols correspond to different features. 
Note that timestamps, while considered during the random walk generation, are not included with the edge features. 
The reason for this choice is to avoid an explosion of tokens if each timestamp were tokenized separately, and allow for inference on unseen future timestamps. 

\textbf{Scheduled Masked Token Prediction.}
Transformer architectures can either be bidirectional or causal. 
Bidirectional models, like BERT, allow tokens at the end of a sequence to affect the attention mechanism for tokens earlier in a sequence. 
As a result, scheduled masked token prediction (SMTP) is used to train them~\cite{smtp}; tokens are masked out of the middle of sequences, and the objective is to guess the missing tokens using the full context of the sequence. 
Causal models do not allow for this.
In language models, this is because we make the assumption that tokens are only affected by those that came before them. 
For example, the next word in a sentence is only affected by those that came before it, not future, unwritten words. 
Causal models use next token prediction (NTP) to enforce this constraint.
While either bidirectional or causal models can be adopted in {\name}, we select BERT~\cite{bert} as the baseline LLM.
As we will later show in Section~\ref{sec:ablation}, the bidirectional attention afforded by using BERT as a base model is more beneficial than a causal model like GPT.

SMTP masks elements in the input sequence, first at a very high rate, then at increasingly lower rates. 
The idea is that at first, the model will learn a general idea of what tokens are associated with the very few input tokens it is allowed to see before gaining more specific knowledge as the training progresses. 

In our implementation, we selected $k$ tokens to perturb according to a polynomial scheduled rate, then of those $k$ tokens, masked 80\% of them, replaced 10\% of them with randomly selected node ids, and did nothing to the remaining 10\%, in following with the original BERT objective function~\cite{bert}. 
Mathematically, the mask rate function is 
\begin{equation}
    \text{MR} = \text{max} \Big\{\bigg( 1 - \bigg(\frac{t}{T_w}\bigg)^2 \bigg)\text{MR}_{\text{fixed}} \hspace{0.5em}, \hspace{0.5em} \text{MR}_{\text{min}}\Big\}
\end{equation}
where $\text{MR}_{\text{fixed}}=0.70$, $\text{MR}_{\text{min}}=0.15$, $t$ is the number of tokens the model has seen during training, and $T_w$ is the number of tokens the model will see before it finishes warming up: in this work, 10\% of the total number of tokens it sees during training. 
We selected the fixed and minimum values of $\text{MR}$ to match those selected by prior work~\cite{graphgpt}. 

The input sequences, now converted to lists of masked tokens, can then be passed through the full model. 
Tokens are used to index the token parameter $\mathbf{T}$, the rows of which are combined with a positional encoding and passed into several transformer encoder layers. 
For the specifics of the architecture, we direct the reader to the original work~\cite{bert}. 
The output of the transformer is a $B \times S \times d$ tensor of token embeddings based on $B$ inputs, with maximal length $S$.
This is passed through a final network that predicts the probability that one of the $d$-dimensional vectors maps to a given token, producing a final $B \times S \times |\mathcal{T}|$ of logits representing the probability that the value at a point in the input sequence was any given token. 

The model is optimized end-to-end, using cross-entropy loss on the logits output by the model. 
In practice, we used the BERT implementation provided by the Hugging Face Transformers python module~\cite{transformers}.

\subsection{Fine-tuning \& Anomaly Detection}
When pretraining has concluded, the model is ready to be fine-tuned for anomalous link detection.
The objective of the fine-tuned models is to predict if the final edge in a given sequence is anomalous or benign.
For a given edge $\langle u,v \rangle$, we sample a random walk from $u$, backwards in time if we are using temporally biased walks, and provide the model with that walk and the node $v$. 

We evaluate two different methods of classifying the likelihood that $v$ is the next node in the random walk: classifier-based and link-prediction-based. 

\textbf{Classifier-based} fine-tuning is similar to traditional LLM fine-tuning. 
Given an edge $\langle u,v \rangle$, and a random walk $\omega_u$ that terminates at $u$, we use $\omega_u || v || \texttt{[CLS]}$ as the model's input. 
Here, \texttt{[CLS]} is a classifier token. 
We then pass the LLM's output embedding of the \texttt{[CLS]} token through an MLP. 
The MLP predicts whether an edge exists between nodes $u$ and $v$. 
We use all edges in the training set as the set of true positives, and randomly generate non-edges to use as the true negatives. 
For each batch, we select an equal number of true edges and negative edges, and pass them into the model. 
The model is then optimized using binary cross-entropy with logits. 

Finally, at test time, we use $1-\sigma \big(f(\langle u,v \rangle)\big)$ as the anomaly score, where $f(\cdot)$ is the fine-tuned GFM, and $\sigma(\cdot)$ is the sigmoid function. 
Edges that the model classifies as negative are considered more anomalous.

This method of fine-tuning works best on graphs that do not have edge features, as negative sampling with edge features is much more difficult. 
For this work, we only use it for graphs without edge features. 
We leave identifying a method of generating negative edges with features as a subject for future work. 

\textbf{Link prediction-based}
anomalous link prediction is similar to the pretraining objective, and we find very few updates are needed to improve performance. 
The graph foundation model has already been optimized to predict masked tokens that appear randomly throughout input sequences. 

For link prediction fine-tuning, given a random walk that terminates at node $u$, we append the edge features $e_v$ and a \texttt{[MASK]} token to the end of the walk. 
The link prediction score for node $v$ is then the product of the likelihoods of the edge features and the node $v$ coming next in the sequence. 
This way, the model predicts the likelihood of an edge existing between a source node and a destination node given the context defined by the random path to the source node. 
This process is repeated for each benign edge in the training set, and the model is again optimized with cross-entropy loss. 
\section{Results}
\label{sec:results}

\subsection{Datasets}
\label{subsec:datasets}
We evaluate our model against three widely used datasets for detecting anomalous activity in a network. 
Table~\ref{tab:meta} contains information describing important features of each dataset after preprocessing. 
We will describe them in more detail below. 

\begin{table}[h]\centering
\caption{Dataset Metadata}\label{tab:meta}
\footnotesize
\begin{tabular}{lrrrrrr}\toprule
&$|\V|$ &$|\E|$ &Anomalies &Days &Edge Features\\\midrule
OpTC~\cite{optc} &976 &1M &674 &8 &0\\
UNSW~\cite{unsw} &50 &2M &321k &27 &180 \\
LANL~\cite{lanl} &15.6k &3M &439 &14 &116 \\
\bottomrule
\end{tabular}
\end{table}

The DARPA \textbf{OpTC} dataset~\cite{optc} contains eight days of activity in a medium-sized network. 
The first five days are benign baseline activity; in the last three days, the network undergoes three different attacks. 
We use the same dataset as that used by~\cite{euler,pikachu,argus}. 
Hosts in the network are represented as nodes, and \texttt{FLOW-START} events are edges between them.
After a host has become compromised on a given attack day, we assume any additional edges eminating from it are also malicious 
This dataset does not contain edge features.

The \textbf{UNSW}-NB15 dataset~\cite{unsw} is built from PCAP records of normal and attack traffic in a small network. 
The dataset contains artifacts from 9 different kinds of attacks, as well as benign data. 
Each event in the dataset comes with 49 features about the protocol, ports, and packets involved. 
We use the source and destination IP addresses as the nodes in our graph representation of the network activity. 
Additionally, we extract several edge features from each interaction. 
We use the most discriminative features as defined by~\cite{unsw_feat_selection}. 
Table~\ref{tab:unsw_feats} lists all of the edge features used. 
In total, 180 unique tokens are required to represent these numeric edge features. 

\begin{table}[h]
    \centering
    \footnotesize
    \caption{Edge features used in UNSW-NB15}
    \label{tab:unsw_feats}
    \begin{tabular}{lp{18em}}
    \toprule 
         \texttt{spkts} & Source to destination packet count\\
         \texttt{sloss} & Source packets retransmitted or dropped\\
         \texttt{dloss} & Destination packets retransmitted or dropped\\
         \texttt{dload} & Destination bits per second\\
         \texttt{ct\_src\_ltm} & Number of connections from the same source address in the last 100 records\\
         \texttt{ct\_srv\_dst} &Number of connections with the same service and same destination in the last 100 records\\
         \texttt{ct\_dst\_src\_ltm}&  Number of connections with the same source and destination IP in the last 100 records\\
    \bottomrule 
    \end{tabular}
    \label{tab:my_label}
\end{table}

The \textbf{LANL} Comprehensive, Multi-Source Cyber-Security Events dataset~\cite{lanl} contains log information of a large computer network over a 58-day period. 
The full dataset includes logging information about processes, flows, and DNS records; however, we only use the authentication and flow logs to build our graph. 
More specifically, as in prior works~\cite{euler,argus}, we only consider NTLM-based authentications. 
We model authentication events as edges in a graph with computers as the nodes. 
Additionally, we use the same method as Argus~\cite{argus} to generate edge features from the network flow data, and only consider the data from the first 14 days as they identified issues in the logs after that point. 

\subsection{Prior works}
We compare our method against four anomaly-based network intrusion detection models. Both Argus~\cite{argus} and Euler~\cite{euler} represent networks as discrete temporal graphs. Their architectures stack a graph neural network (GNN) on a recurrent neural network (RNN) to process the dynamics of node embeddings over time. Euler uses a loss function to optimize log loss, while Argus optimizes for AP directly, using a specialized loss function~\cite{aploss}. Argus can also encode edge features using edge-conditioned convolution~\cite{edgeconv} rather than a GNN for the final graph embedding layer. We select these methods as a representative sample of the best GNN-based link prediction methods for network anomaly detection. 

We also compare to Pikachu~\cite{pikachu}, a temporally-biased random walk-based method, and N2V~\cite{bowman2020}, a static random walk-based method. 
Both methods generate node embeddings using Node2Vec to perform link prediction. 
N2V~\cite{bowman2020} uses embeddings built from random walks through the graph when viewed as a static entity. 
Pairs of embeddings are then passed through a module to perform anomaly detection, optimizing for log loss. 

Pikachu advances this work by representing the network instead as a temporal graph. 
It embeds nodes based on temporally-biased random walks before passing them through a more advanced anomaly detection module. 
We select these two methods to contrast our random-walk based method, which aims to advance Node2Vec into the age of transformers, with analogous approaches that use traditional skip-gram embedding of their random walks. 

\subsection{Setup}

\begin{table}[t]
    \centering
    \footnotesize 
    \begin{tabular}{lcccc}
    \toprule 
         &Parameters	&RW-based &Temporal &Edge Features\\
    \midrule 
        N2V~\cite{bowman2020} &1.72M &\checkmark \\
        Pikachu~\cite{pikachu} &3.31M &\checkmark &\checkmark \\
        Euler~\cite{euler} &2.05M & &\checkmark	\\
        Argus~\cite{argus} &2.31M & &\checkmark &\checkmark \\
        {\name} &2.57M &\checkmark &\checkmark &\checkmark \\
    \bottomrule
    \end{tabular}
    \caption{Model comparison. Parameter counts are those used for the LANL dataset.}
    \label{tab:modelspecs}
\end{table}

For all reported results, we use the BERT tiny architecture (2 layers, 2 attention heads, 128-dimensional embeddings, and 512-dimensional hidden layers). 
We selected this smaller model to have a comparable number of parameters to the prior works and ensure fair comparison.\footnote{BERT-mini uses 7M parameters, BERT-small increases to 21M}
However, it also demonstrates the potential power of these models.
As we will show, we are able to achieve state-of-the-art precision on every dataset. 
In the NLP space, it has been shown that models following this architecture have associated scaling laws~\cite{llm_scaling} where performance is directly related to the size of the model. 
We show in the supplemental material that this also occurs for GFMs, where we achieve even higher scores when models have 10x more parameters. 

During pretraining, we sample one walk of length 64 tokens per node. 
As is common with LLMs, we measure progress in terms of how many tokens the models have learned from rather than epochs. 
Models trained on the OpTC and UNSW datasets are trained for $10^8$ tokens, and the LANL model is trained for $10^9$ tokens. 
Every $10^7$ tokens, we evaluate the models on the validation set, using the link prediction-based fine-tuning algorithm on walks of length 1. 
This is to say, we provide the models with a source node, and edge features, and use one minus the likelihood of the destination node as its anomaly score. 
The pretrained model with the highest validation score using this method is saved and used for fine-tuning. 

During the fine-tuning stage, we define an epoch as a learning step over every available training edge. 
Models are fine-tuned for 5 epochs, and evaluated at the end of every epoch. 
The score on the test edges for the epoch with the highest AUC on the validation data is what is reported. 
During pretraining and fine-tuning, models use a scheduled learning rate with the AdamW optimizer~\cite{adamw}. 
We use the same hyperparameters as GraphGPT~\cite{graphgpt} for these rates, as well as all hyperparameters other than epochs. 
The specific values of all hyperparameters are detailed in Table~\ref{tab:hyperparams}.
All models are trained on a single Tesla V100 GPU with 32GB of VRAM.

\begin{table}[!htp]
\centering
\scriptsize
\caption{Model hyperparameters}\label{tab:hyperparams}
\begin{tabular}{lcc}\toprule
&Pretraining &Fine-tuning \\\midrule
Batch size &1024 &1024 \\
Total &$10^8$ , $10^9$ Tokens &5 epochs \\
Warmup tokens &Total / 10 &Total / 3.33 \\
LR Scheduler &Linear decay &Cosine decay \\
Max LR &3e-4 &3e-5 \\
Adam betas &[0.9,0.95] &[0.9,0.99] \\
Adam eps &1e-8 &1e-10 \\
Max grad norm &5 &1 \\
Weight decay &0.1 &0 \\
\bottomrule
\end{tabular}
\end{table}

\subsection{Anomalous Network Activity Detection}
To measure the importance of temporal bias that has been claimed across several prior works~\cite{pikachu,euler,argus} we evaluate our system trained on temporally biased random walks ({\name}$_\text{T}$) as well as a system trained on random walks through the static graph ({\name}$_\text{S}$) on the OpTC, UNSW, and LANL datasets. 
Each dataset is partitioned such that 80\% of the benign data is used for training, 10\% is used for validation, and the remaining 10\%, as well as the anomalous data, is used for testing. 

We note that many prior works use a temporal data split, training exclusively on data from before the time of the attack. 
This split was infeasible in this case for a number of reasons: in the UNSW dataset, the network is under constant attack from the beginning. 
In the LANL dataset, many hosts do not appear in the logs until during or after the attack. 
Our model's inability to inductively represent unseen hosts is a limitation that will be a subject of future work, but it is a limitation shared by all random-walk-based anomaly detection models. 
To ensure proper comparison, we independently evaluated each model using the same data splits across each dataset. 
Each model is evaluated 10 times, and the reported results are the average across all of the runs using the model weights with the highest AUC on the validation set during training. 
Due to resource constraints, the {\name} models are only pretrained once, and the average of 10 rounds of fine-tuning is reported. 
Table~\ref{tab:results} shows the results of the best performing models for each dataset. 

\begin{table}[!htp]\centering
\caption{Results of Anomalous Link Prediction Tests. Best results in \textbf{bold}, second-best results \underline{underlined}.}\label{tab:results}
\scriptsize
\begin{tabular}{lrrrrrrr}
\toprule
&\multicolumn{2}{c}{OpTC} &\multicolumn{2}{c}{UNSW} &\multicolumn{2}{c}{LANL} \\\cmidrule{2-7}
&AUC &AP &AUC &AP &AUC &AP \\\midrule
N2V &0.9420 &0.3519 & 0.3897 &0.6814 &0.9495 &0.0605 \\
Pikachu &0.9707 &0.3879 &0.9383 &0.9190 &0.9709 &0.1428 \\
Euler &\textbf{0.9840} &0.6423 &0.7162 &0.7918 &0.9719 &0.0627 \\
Argus &0.8885 &0.7847 &0.9375 &0.9181 &0.9817 &0.2279 \\\midrule
$\text{CyberGFM}_{\text{S}}$ &0.9739 &\textbf{0.8981} &\underline{0.9931} &\underline{0.9874} &\textbf{0.9994} &\textbf{0.7600} \\
$\text{CyberGFM}_{\text{T}}$ &\underline{0.9759} &\underline{0.8870} &\textbf{0.9942} &\textbf{0.9929} &\underline{0.9983} &\underline{0.5702} \\
\bottomrule
\end{tabular}
\end{table}

We observe that across all datasets, {\name} has the best precision. 
The gain in precision on the LANL dataset is especially notable; the LANL dataset is notoriously difficult, but our method more than doubles the best precision on this dataset across the prior works. 
While in one instance, Euler has a higher AUC score than our method on the OpTC dataset, it also comes with a 0.23 lower AP score, which is far more important for tasks with such imbalanced data. 
We also observe that, surprisingly, the static {\name} model has better performance overall. 
With the exception of the UNSW dataset, and the AUC score on the OpTC dataset, representing the graph as static rather than temporal yielded higher metrics. 
One reason for this may be the length of possible random walks in static vs dynamic graphs. 
Nodes that are part of connected components can generate walks of arbitrary length in static graphs, while temporally biased walks are necessarily constrained to a maximum possible length, as each edge in the walk must have occurred later in time than the previous one. 
Empirically, we observed that the average walk length in static graphs was close to the maximum allowed walk length, while in temporal graphs, it was often half. 

\subsection{Ablation Studies}
\label{sec:ablation}

\textbf{Fine-tuning}: 
An important stage of our proposed pipeline is the fine-tuning step. 
While, especially for the link prediction-based fine-tuning methods, it may appear that the unsupervised training objective for {\name} is sufficient to produce an anomaly detection model, we found that the additional fine-tuning step is critical to improving model precision. 
During the SMTP training step, the models learn to fill in missing tokens wherever they appear in the sequence, but during fine-tuning, they specifically learn to predict tokens at the end of a sequence. 
Also, importantly, the sequences are fixed-length and in some cases much shorter than what the models have seen during training. 

\begin{figure}[t]
    \centering
    \includegraphics[width=0.9\linewidth]{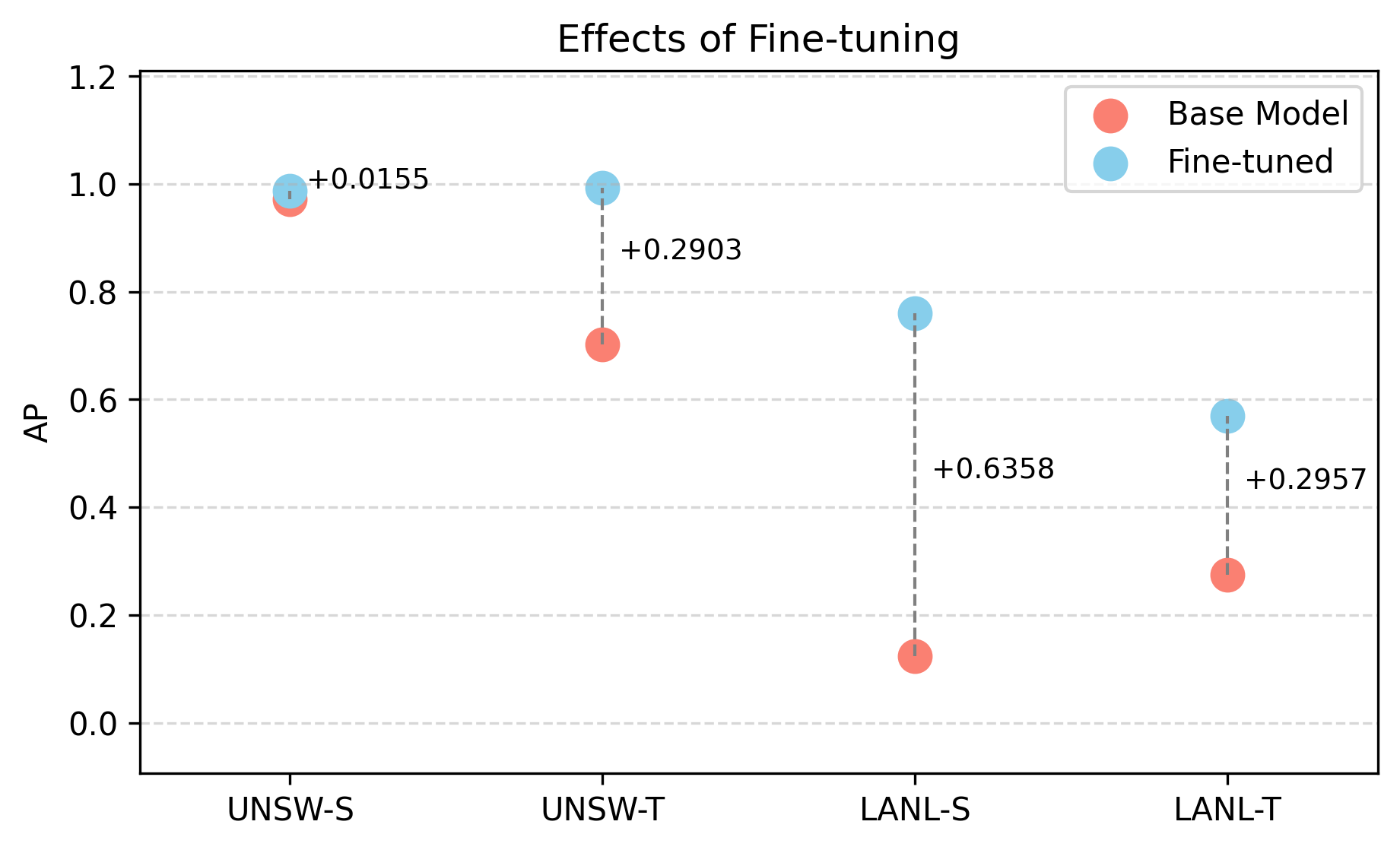}
    \caption{Change in AP score before and after fine-tuning the {\name} models}
    \label{fig:ft_effects}
\end{figure}

These distinctions may seem unimportant, but as illustrated in Figure~\ref{fig:ft_effects}, fine-tuning the models results in significant increases in their average precision. 
Note that we cannot directly compare the increase afforded by fine-tuning in the case of the OpTC dataset, as we used CLS fine-tuning for this dataset because it had no edge features. 
However, for the remaining two datasets, the link prediction task is directly applicable to both the base and fine-tuned models. 
We observe that the base models on their own can, in most cases, outperform prior work; however, the fine-tuning process brings significant improvements. 
For example, in the case of the LANL dataset, the pretrained temporal model has an AP of 0.2745. This score is already higher than the prior works we tested. 
After fine-tuning, the AP increases by as much as 0.63 in the most extreme case. 
The UNSW dataset benefited the least from fine-tuning, which is likely explained by its simplicity: there are only 50 nodes, and 180 unique edge features, compared to the 15.6k nodes, and 116 edge features present in LANL.

\begin{figure}[t]
    \centering
    \includegraphics[width=\linewidth]{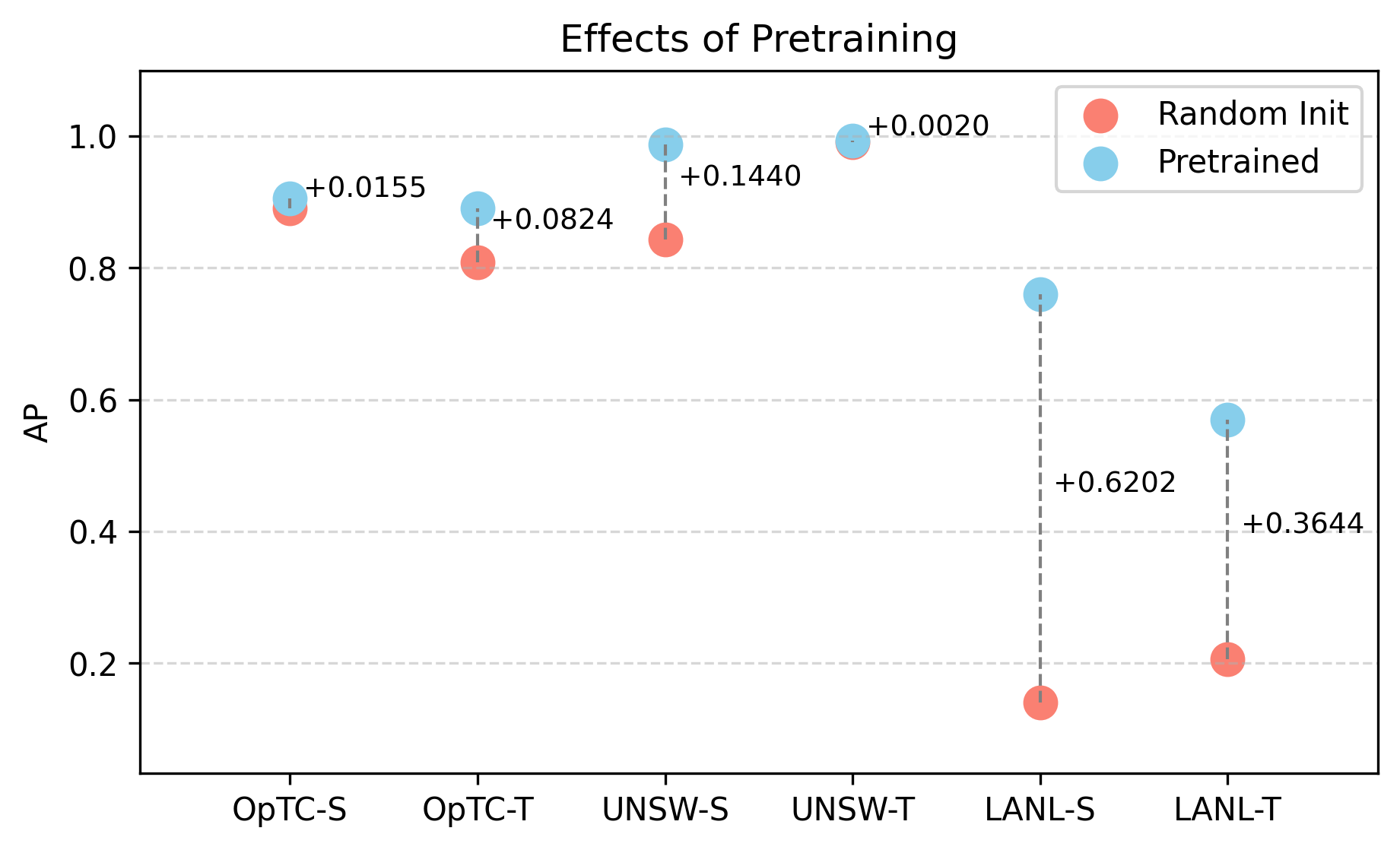}
    \caption{Effect of pretraining before the finetuning stage}
    \label{fig:pretrain_effects}
\end{figure}

\textbf{Pretraining}: Given the impressive boosts in score we observed from the fine-tuning step, it is natural to wonder if the pretraining step is necessary at all. 
Unlike traditional text foundation models, our downstream task is very similar to the unsupervised pretraining task. 
Additionally, the fine-tuning step is not strictly supervised; as we are interested in anomaly-based intrusion detection, we cannot assume that we have \textit{any} labeled data, and must adapt another unsupervised approach during fine-tuning. 
Nonetheless, it is theorized that pretraining plays a regularizing role in the overall training process~\cite{why_does_pretraining_work}, and it has been empirically demonstrated to improve evaluation criteria, as well as speed up convergence~\cite{unsupervised_pretrain}. 
To measure the utility of pretraining in our pipeline, we evaluate models given random initializations before the fine-tuning process to models that were pretrained. 
The results of these experiments are shown in Figure~\ref{fig:pretrain_effects}

\begin{figure}[t]
    \centering
    \includegraphics[width=\linewidth]{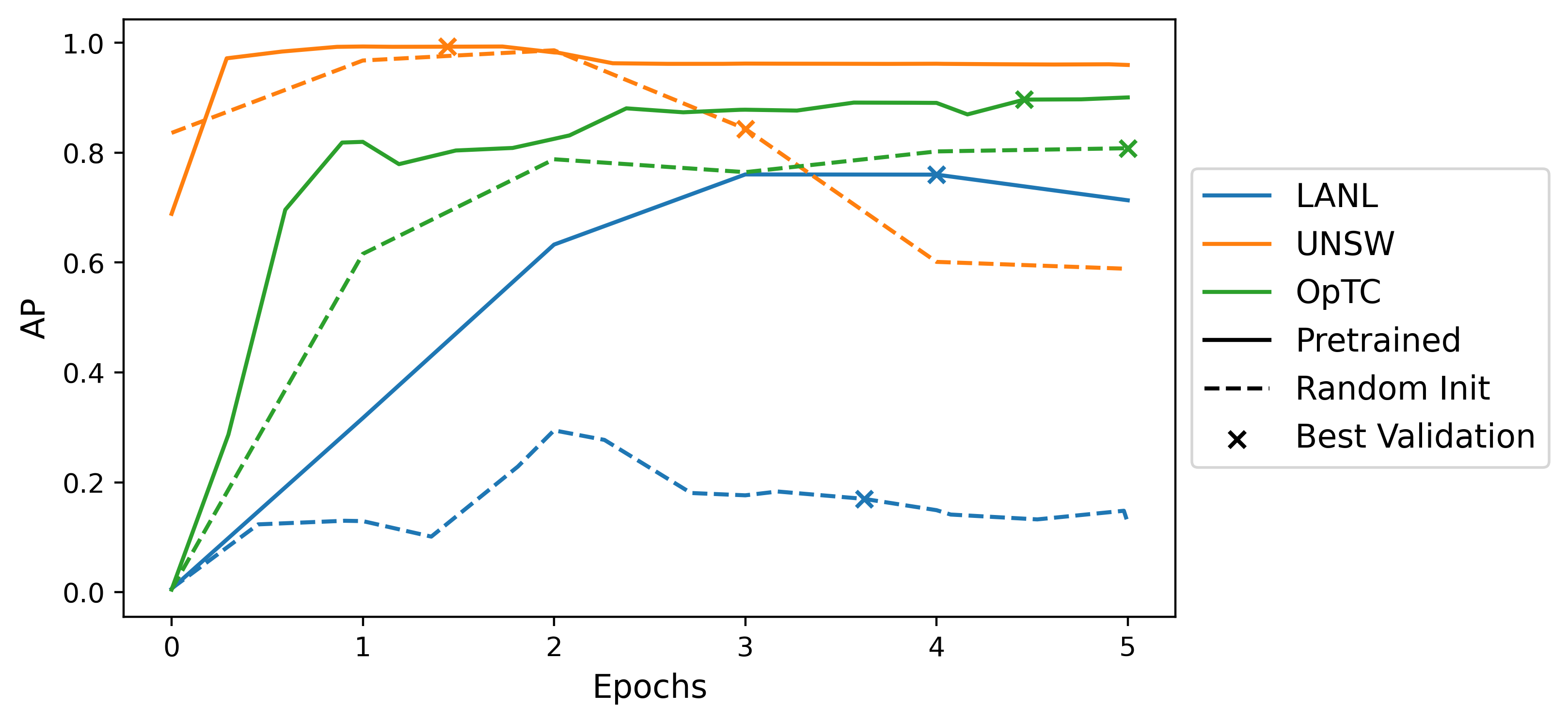}
    \caption{AP over time for pretrained and randomly initialized models}
    \label{fig:pretrain_convergence}
\end{figure}

In all cases, the pretrained models learned functions with greater average precision. 
Though for the datasets with fewer unique tokens, the increase is less pronounced.
However, the raw numbers only tell part of the story. 
During the fine-tuning process, we observed that the pretrained models exhibited much higher stability in addition to the better average precision they were able to attain. 
Figure~\ref{fig:pretrain_convergence} shows the trajectory of the AP as the models were fine-tuned. 
We note that in both the UNSW and LANL datasets, the evaluation begins decreasing after three epochs instead of remaining high. 
It is also noteworthy that the validation scores of the randomly initiated model weights did not accurately reflect the test scores; when test scores decreased, the validation scores continued increasing. 
This is shown in the chart by the ``x" marks: each mark is the update with the highest validation score. 
The disconnect between the validation and testing scores suggests that the randomly initialized models are overfitting to the data~\cite{overfitting}. 
These results support the theory that pretraining has a regularization effect and is a valuable step in the pipeline.

\textbf{Walk lengths}: During fine-tuning, models are given fixed-length random walks that terminate at the source node of a given edge. 
The length of these random walks is a tunable parameter that deserves closer inspection. 
Models trained with longer random walks are given more context about the neighborhood surrounding the source node, but longer walks can also produce irrelevant noise. 
Additionally, longer walks require more memory, which requires smaller minibatch sizes, slowing down the training process.

\begin{figure*}[t]
    \centering
    \begin{subfigure}{0.32\textwidth}
        \centering
        \includegraphics[width=\linewidth]{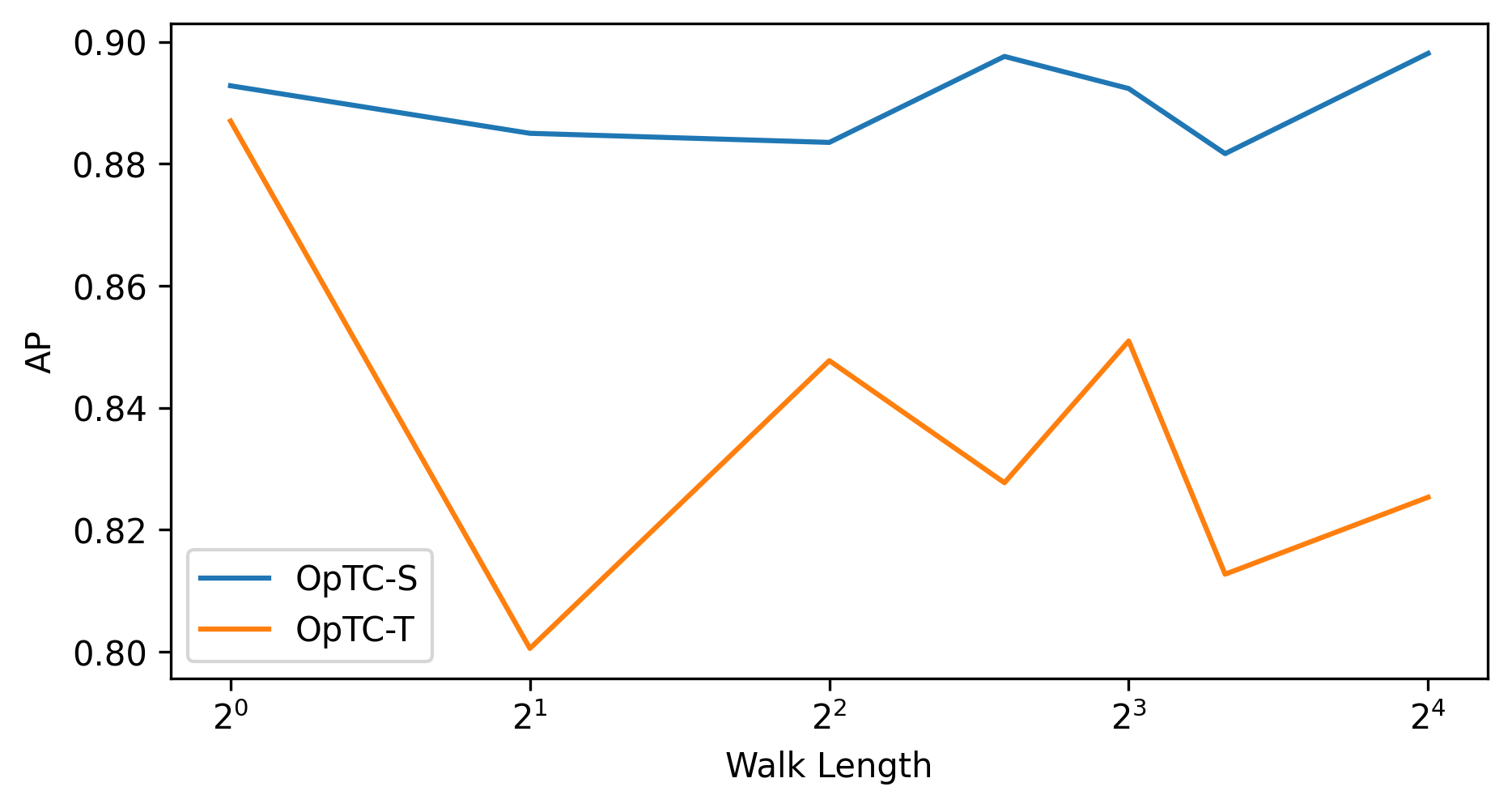}
        \caption{OpTC}
        \label{fig:wl_optc}
    \end{subfigure}%
    \hfill%
    \begin{subfigure}{0.32\textwidth}
        \centering
        \includegraphics[width=\linewidth]{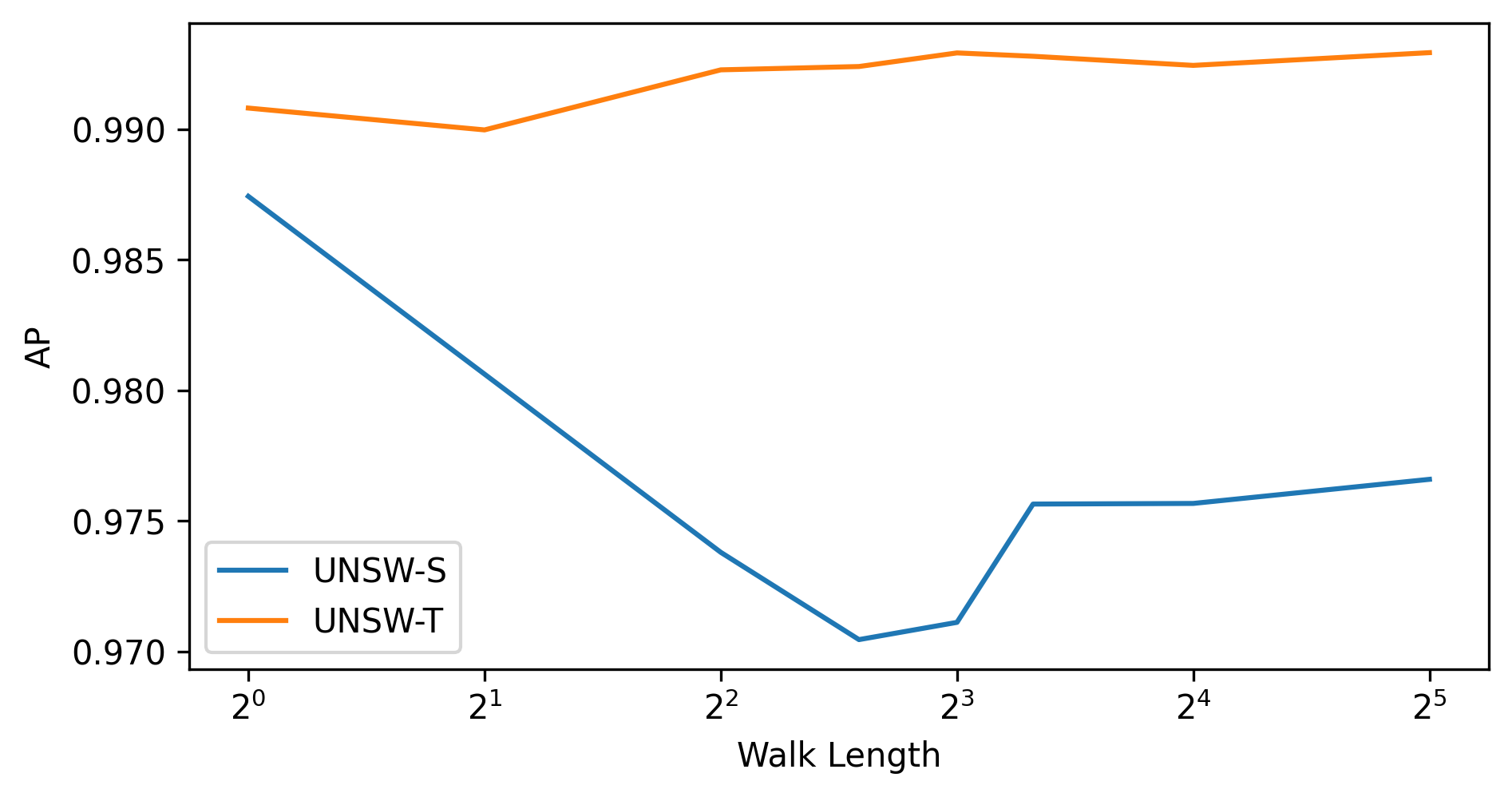}
        \caption{UNSW}
        \label{fig:wl_unsw}
    \end{subfigure}%
    \hfill%
    \begin{subfigure}{0.32\textwidth}
        \centering
        \includegraphics[width=\linewidth]{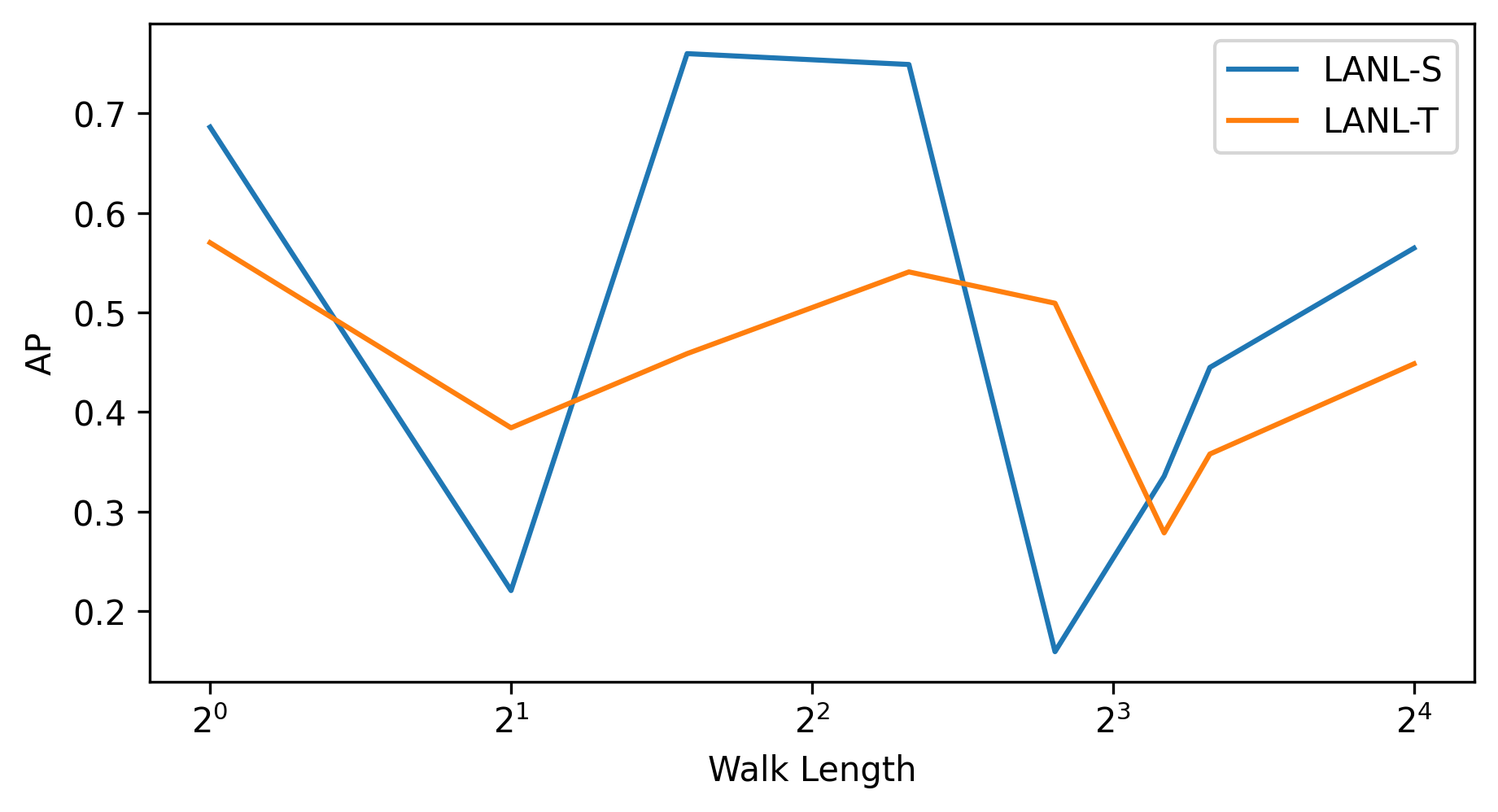}
        \caption{LANL}
        \label{fig:wl_lanl}
    \end{subfigure}
    \caption{Ablation study on how the walk length used during fine-tuning affects average precision. Note that AP ranges on Y-axis are different for three datasets.}
    \label{fig:wl_ablation}
\end{figure*}

In Figure~\ref{fig:wl_ablation}, we plot the AP score attained by fine-tuning with various walk lengths across each dataset. 
One surprising result is that walk lengths of 1, that is, models trained with only the source node and edge features as input, have surprisingly high performance. 
In the case of the UNSW dataset, the static model had its best performance with a walk length of 1. 
This is likely due to the simplicity of the dataset. 
Many prior works~\cite{unsw_benchmark1, unsw_benchmark2} have achieved near-perfect scores on this dataset using supervised approaches that only consider the features of packets, ignoring the relational aspect of the dataset. 
The relationships between hosts may be less important in this case than the edge features, highlighting the importance of models' capability to consider them. 

Another surprising result is the sensitivity of the LANL models to walk length. 
While in general, the trend is that longer walks result in better scores, the region with walk lengths 2-4 forms a local maximum. 
There is also an outlier when walks have a length of 9, highlighting the importance of a full parameter sweep. 

Models that use temporally biased random walks are less sensitive to their length. 
The temporal biasing could be a regularizing factor that makes walks more regular and decreases the model's variance. 
Given how little labeled data is available to validate anomaly detection models, in practice this regularization is likely a desirable feature of these models. 
Users of this approach may be willing to trade off some precision to have a more robust model.  

\begin{figure}
    \centering
    \includegraphics[width=\linewidth]{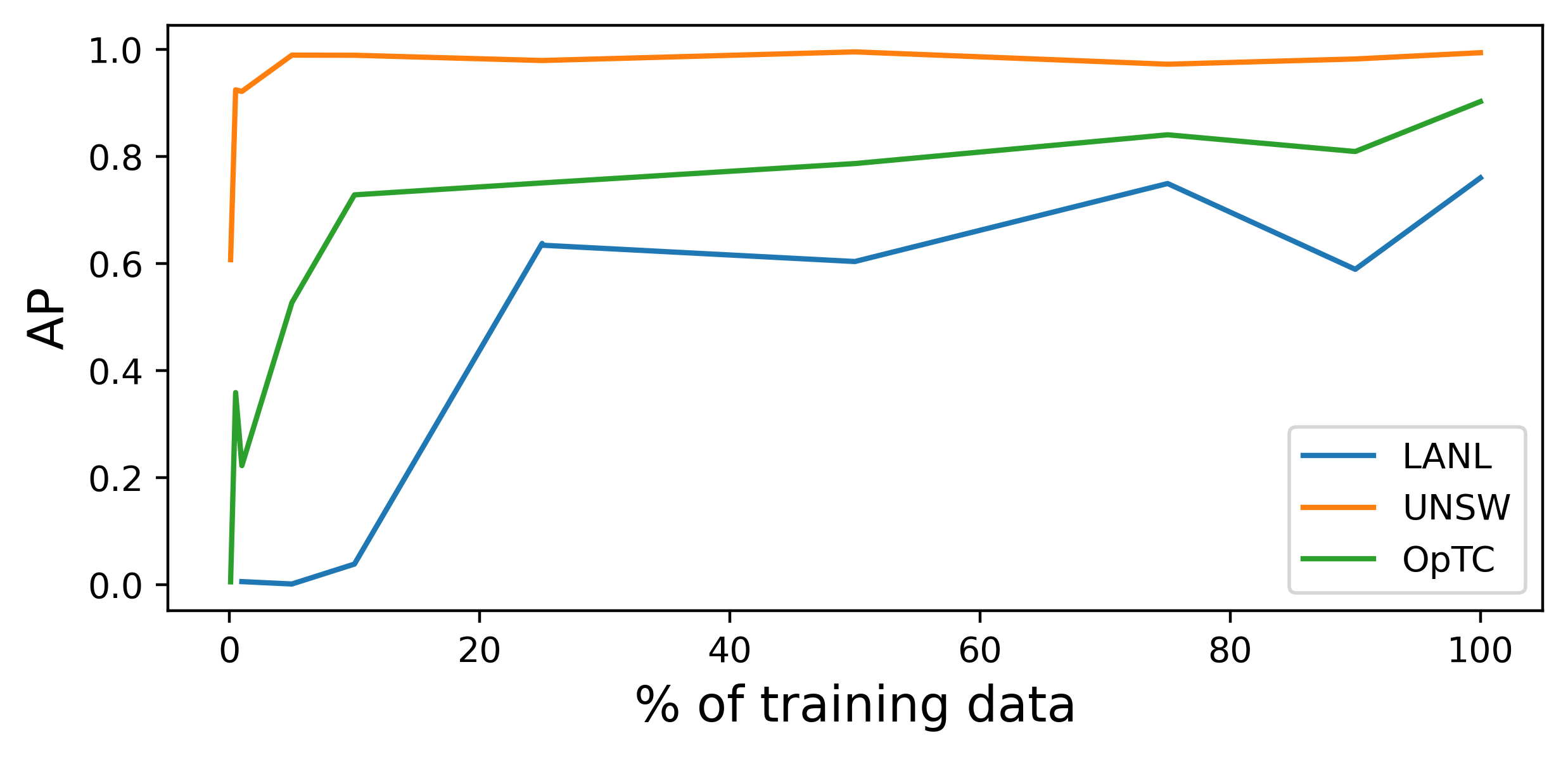}
    \caption{Effect of using less training data for end-to-end pretraining and fine-tuning}
    \label{fig:tr_size}
\end{figure}

\textbf{Training set size}:
Though it is always useful to have a large body of training data to work with, it is not necessarily a realistic assumption that we will have it. 
To measure the robustness of our approach to data availability, we perform additional evaluations where we partition the dataset into progressively smaller subsets, and retrain the model from end to end. 
Figure~\ref{fig:tr_size} shows how AP score increases as additional data is added to the training set. 
Unsurprisingly, the models with the most training data had the best performances. 
However, as long as at least 25\% of the training data is present, models continue to perform well. 
The dropoff in quality does not appear until only 10\% of data is used in the case of LANL, and 5\% in the case of the smaller datasets. 
The model maintains state-of-the-art performance on the LANL dataset even with 75\% of the training data missing. 

\textbf{Bidirectional vs causal model}: We conduct an additional experiment using GPT, a causal, NTP model, to compare with a BERT model. 
Table~\ref{tab:bert_v_gpt} shows the results of these experiments. 
We observe that while GPT still achieves results better than the prior works we tested, it performs worse than the BERT model. 
This is likely because a node in the context of a random walk is indeed affected by the nodes that come after. 
For example, consider a random walk $\{v_0, v_1, v_2, v_3, v_4\}$. 
With BERT as the underlying FM, node $v_2$ is encoded with context from both $v_0$ and $v_4$. 
It is valuable information that $v_2$ is two hops from both of those nodes. 
However, using GPT as the underlying FM, no information about $v_3$ or $v_4$ is taken into account in the representation of $v_2$. 
As the empirical results show, this loss of information will affect the overall model significantly, especially for larger datasets like LANL. 

\begin{table}[t]
    \centering
    \scriptsize
    \begin{tabular}{llrrrrrrr}\toprule
    & &\multicolumn{2}{c}{OpTC} &\multicolumn{2}{c}{UNSW} &\multicolumn{2}{c}{LANL} \\\cmidrule{3-8}
    & &AUC &AP &AUC &AP &AUC &AP \\\midrule
    \multirow{2}{*}{BERT} &S &0.9739 &0.8981 &0.9931 &0.9874 &0.9994 &0.7600 \\
    &T &0.9759 &0.8870 &0.9942 &0.9929 &0.9983 &0.5702 \\\midrule
    \multirow{2}{*}{GPT} &S &0.9707 &0.7917 &0.9792 &0.9592 &0.9433	&0.2636 \\
    &T &0.9651 &0.8791 &0.9836 &0.9701 &0.9743 &0.2733 \\
    \bottomrule
    \end{tabular}
    \caption{Results of using different FMs for {\name}. S stands for random walks on a static graph and T for temporally biased random walks.}
    \label{tab:bert_v_gpt}
\end{table}

One interesting thing to note is that with causal models, training with temporally biased random walks results in higher-performing models. 
This is likely because the model is learning the causal relations within random walks, rather than purely structural relations, as it would with the underlying BERT models. 
This result makes sense intuitively, but as we have demonstrated empirically, structural information about a network appears more valuable than temporal information, even when the underlying model is causal. 
Thus, we conclude that temporal information is not as important as prior works assumed it was. 
Perhaps more expressive means of representing graph structure are all that is needed to advance link prediction for lateral movement detection.
\begin{figure*}
    \centering
    \begin{subfigure}{0.32\textwidth}
        \includegraphics[width=\linewidth]{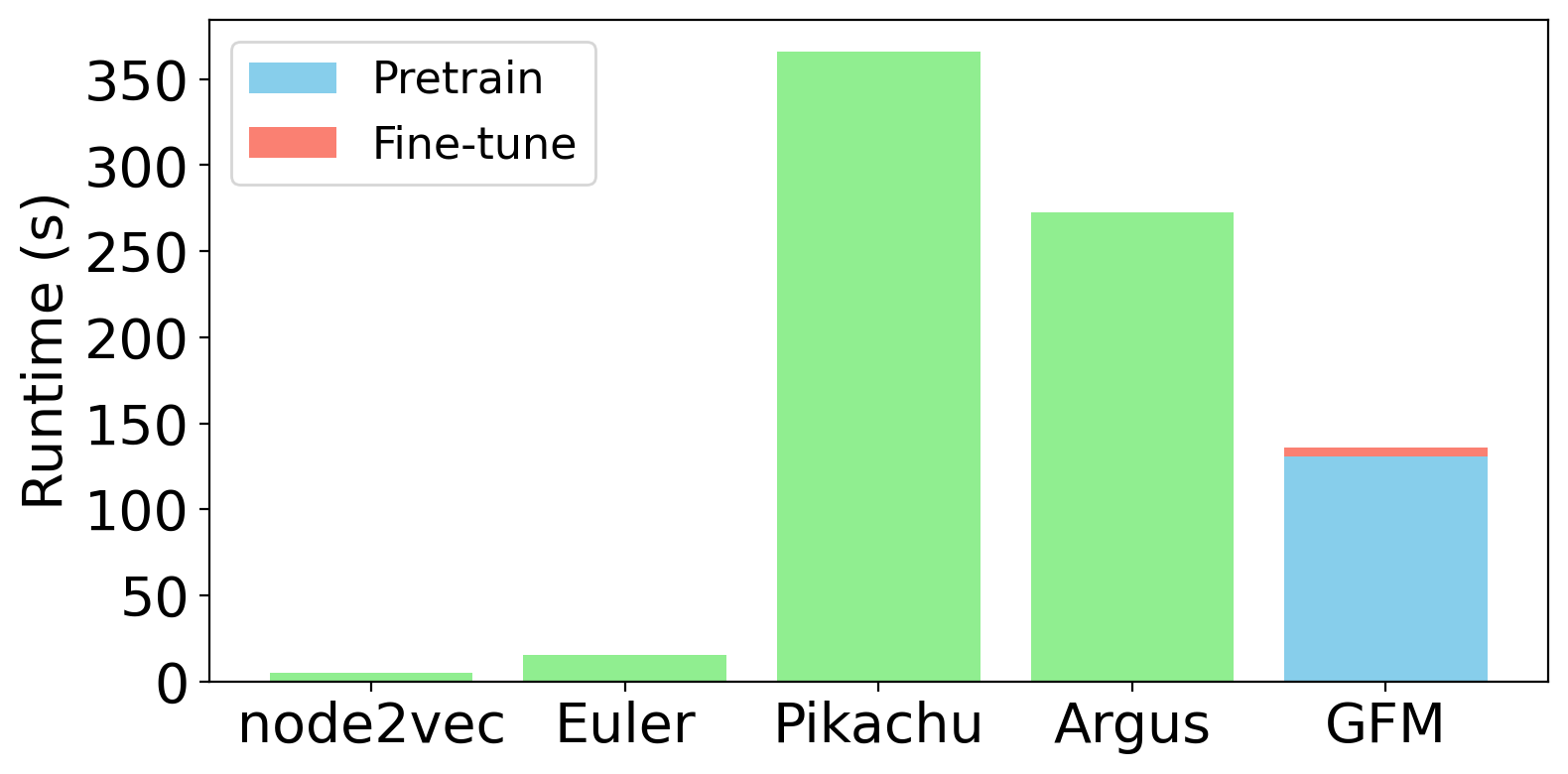}
        \subcaption{OpTC}
        \label{fig:placeholder}    
    \end{subfigure}%
    \begin{subfigure}{0.32\textwidth}
        \includegraphics[width=\linewidth]{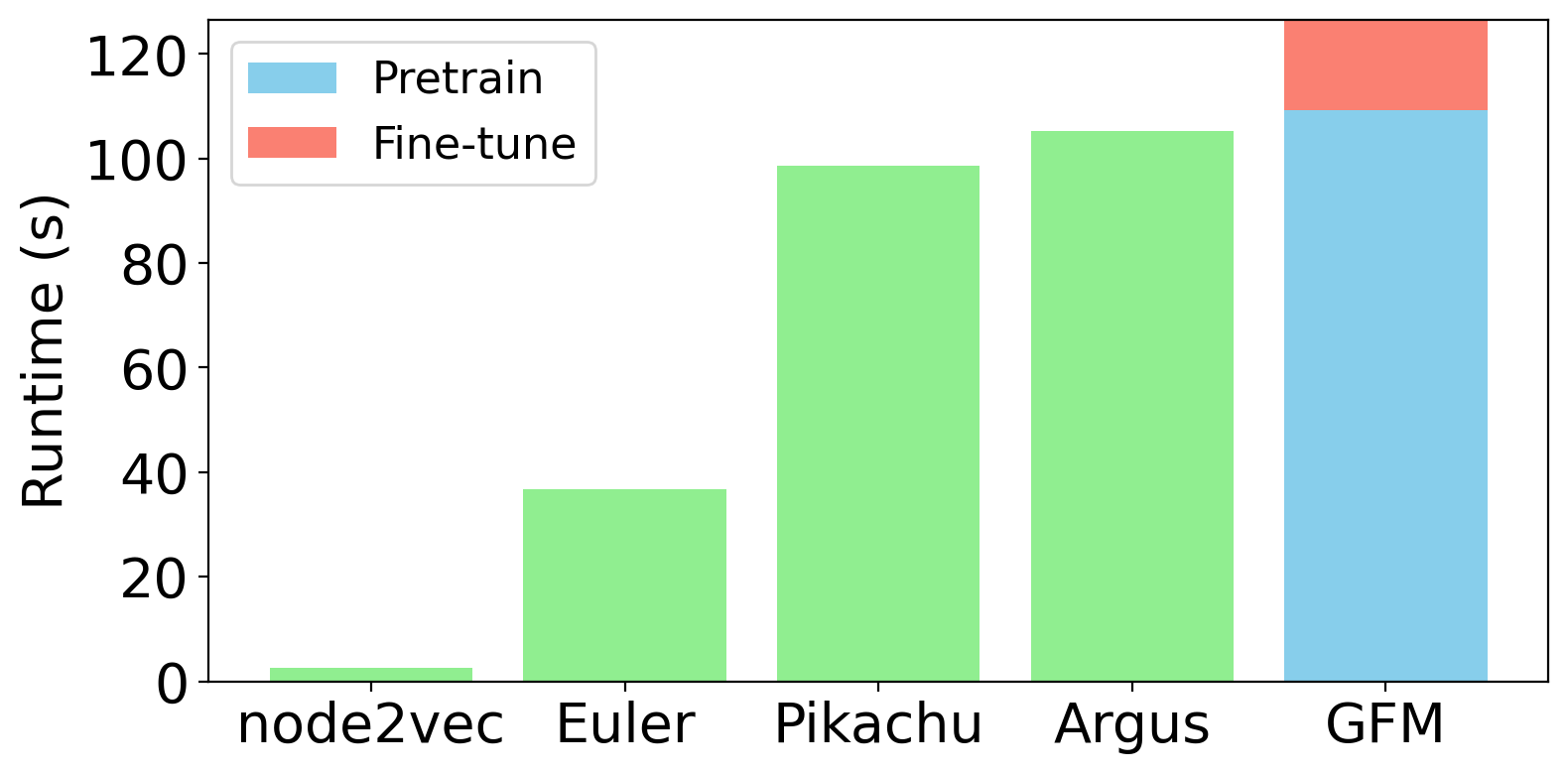}
        \subcaption{UNSW}
        \label{fig:placeholder}    
    \end{subfigure}%
    \begin{subfigure}{0.32\textwidth}
        \includegraphics[width=\linewidth]{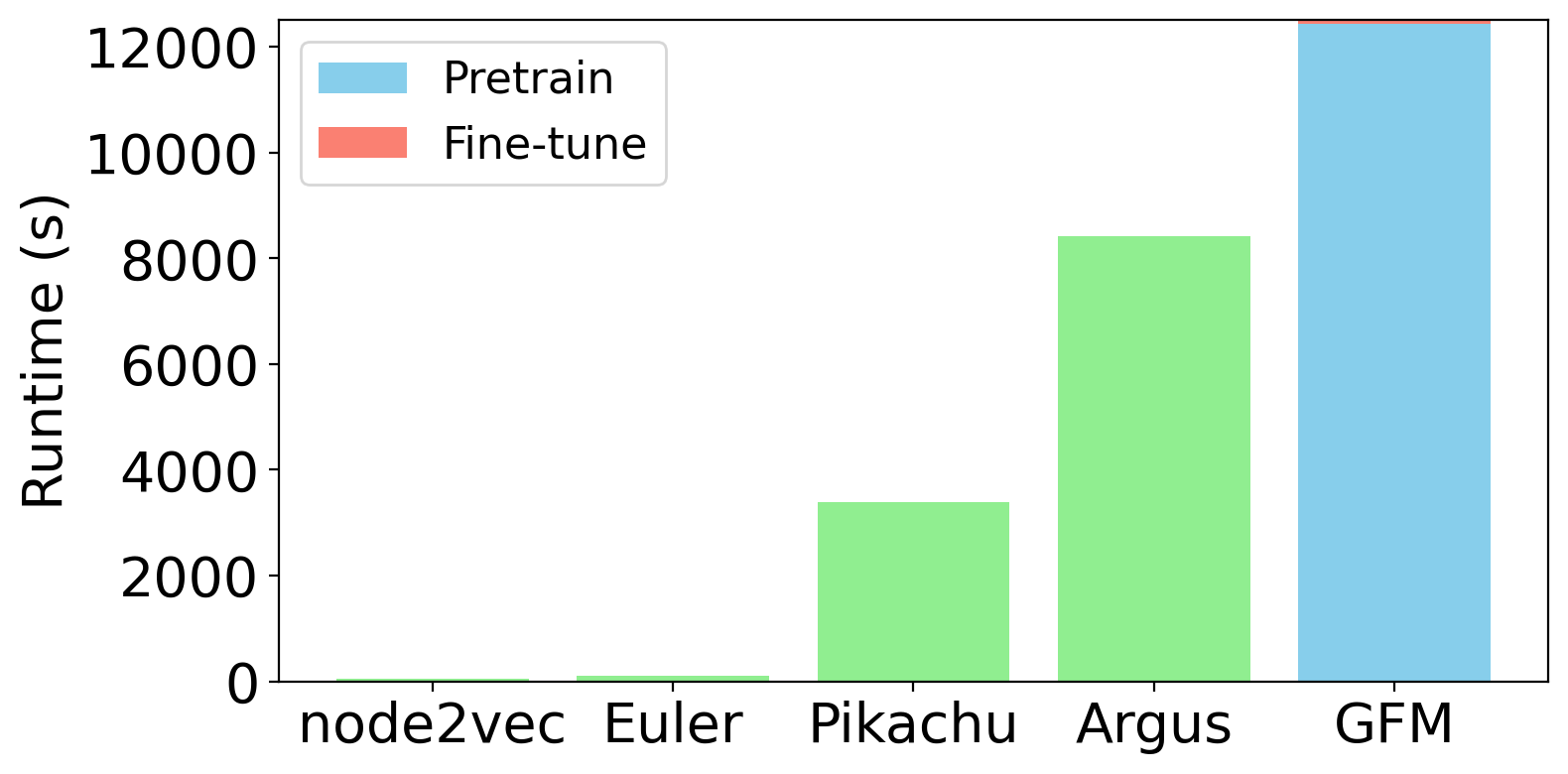}
        \subcaption{LANL}
        \label{fig:placeholder}    
    \end{subfigure}
    \caption{Complete traning time comparison for each model.}
    \label{fig:speeds}
\end{figure*}

\section{Efficiency Study}
The design of the {\name} system remains efficient despite its size for two primary reasons: it is random-walk based, meaning its complexity grows as a function of nodes rather than edges, and it is built from transformers, which are specifically optimized for GPU calculations. 
These two attributes allow it to scale to graphs with millions of edges as well as, or better than prior works in most cases. 
Figure~\ref{fig:speeds} illustrates the complete runtime required for {\name} to train on each dataset evaluated. 
While Euler and Node2Vec are consistently the fastest algorithms, they are also the models with the poorest evaluation scores. 
Pikachu, Argus, and our method are the top three methods by a wide margin in all tests, so we focus on quantifying how much the extra time it takes to train these models is reflected in their evaluation improvements. 

Importantly, our model is trained for a fixed number of steps. 
As is common with language models, we measure progress in terms of how many tokens the models have seen rather than epochs. 
As such, in both the OpTC and UNSW datasets, our model takes a fixed 100s to complete pretraining.
In both cases, the system continually samples random walks until it has viewed $10^8$ tokens, then moves on to fine-tuning for a fixed 5 epochs. 
Pikachu generates temporally biased Node2Vec embeddings for each time window present in the training data, and Argus generates node embeddings for each time window. 
Both methods have an asymptotic lower bound of $\Omega(T|\V|)$ per epoch, where $T$ denotes the number of time windows or snapshots the models train on. 
The OpTC dataset has values $|\V|=1000$ and $T=193$ ($T|\V|=193,000$), while UNSW has $|\V|=50$, $T=676$ ($T|\V|=33,800$), which explains difference in efficiency across these two datasets. 

On the larger LANL dataset, our model again trains for a fixed amount of time; now for $10^9$ tokens. 
This extra training time comes at considerable expense, resulting in a 12,000s training time. 
This is about 4,000s ($\sim$1h) longer than Argus, and 8,000s ($\sim$2.5h) longer than Pikachu. 
This is significant, but potentially worth the gain in precision afforded by our model. 
However, if time is an important consideration, we can simply lower the maximum tokens that the model sees before training is considered finished. 

\begin{figure}
    \centering
    \includegraphics[width=\linewidth]{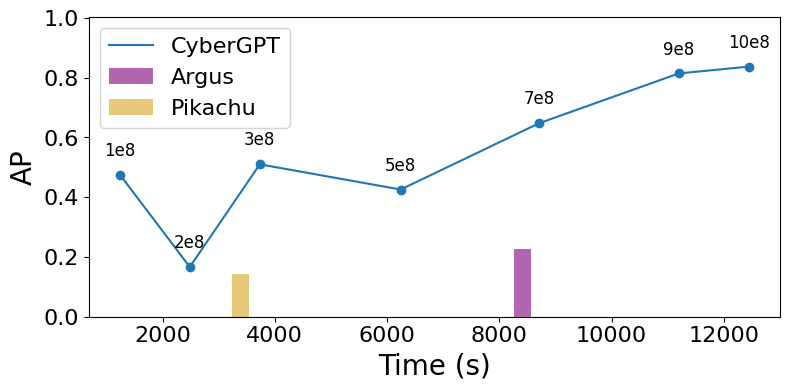}
    \caption{Wallclock time vs. average precision on the LANL dataset. The dots on the line represent the number of pretraining tokens for {\name}}
    \label{fig:time_vs_ap}
\end{figure}

We conduct an additional study on how varying the number of tokens {\name} uses to pretrain affects its final score on the LANL dataset. Figure~\ref{fig:time_vs_ap} shows that even when the set of training tokens is constrained to $10^8$, and runtime is 1,200s, the AP is still higher than all prior works. 
Increasing the training tokens to $10^9$ more than doubles the evaluation scores, though it also requires 10x more wall-clock time to train.
As we are primarily interested in the most precise lateral movement detector possible, we believe this is an acceptable tradeoff, especially given that the wall-clock times in question are 3 hours vs 20 minutes. 
Further, the trend of Figure~\ref{fig:time_vs_ap} suggests that extending the training time beyond $10^9$ tokens may provide additional benefits, which we plan to explore as future work. 

\section{Discussion}
\label{sec:discussion}

\textbf{Pretraining}:
We have demonstrated that pretraining the GFM results in greater stability and less variance during fine-tuning, but the results of the efficiency study raise questions about its value. 
In the smaller datasets, skipping this step results in a 5x speedup during training, and very little difference in the final precision of the model. 
This is a surprising result, but it is likely explained by size of the datasets that were evaluated. 
Recent studies~\cite{pretrain_v_from_scratch} have shown that in small datasets, models trained ``from scratch", that is, without pretraining, can outperform pretrained models as the smaller dataset favors memorization and bias. 
While we did not observe this trend, we did observe smaller benefits from pretraining than expected on datasets represented in fewer than 1k tokens. 
Continuing to quantify the benefits of pretraining on larger datasets with larger models will be a subject of our future work. 

\textbf{Temporal Bias}:
Another surprising result was how little benefit temporal bias provided. 
While previous works all take the importance of temporal dynamics as a given, we found that models without RNNs sometimes outperformed models that used them. 
Our results showed that temporal bias actually affected the results negatively in some cases. 
This is likely due to the smaller walk lengths allowed by temporally constrained walks. 
We only considered sampling walks using a uniform distribution for all allowed future edges; it is possible to constrain this using an exponential distribution, ensuring most edges in the paths are ``nearby" in time. 
This would ensure longer walks, but it is unclear if this would lead to any performance benefits. 

\textbf{Limitations \& Future Work}:
Like all random walk-based approaches, including prior works~\cite{bowman2020,pikachu}, a major limitation of our approach is that it is non-inductive. 
This means it will only be able to evaluate traffic between nodes it has seen before. 
However, there are ways around this. 
Given more robust node features, node IDs can be forgone entirely, and raw features could be tokenized and used as inputs to the random walks. 
Unlike purely random-walk-based approaches, which build on Word2Vec, our approach could be made to be inductive, given descriptive input features about each host.
As these features did not exist in these datasets, all of the models were technically non-inductive. 
Even Argus and Euler, while inductive by design, use node IDs as input features for these datasets, and would also require retraining given new hosts added to the network. 

Another limitation with this approach is scalability. 
Though it was not a problem for the datasets we used, as more hosts are added to the network, if we continue to represent them using node IDs, eventually the embedding table would grow too large to be usable. 
Prior works on Graph Foundation Models, such as GraphGPT~\cite{graphgpt} solve this by splitting node IDs into multiple tokens, but we again stress the importance of inductivity. 
This problem would be completely alleviated in the presence of useful node features to use as inputs. 

Finally, we concede that inference using our method is somewhat inefficient. 
To calculate a given edge's anomaly score requires a random walk through the graph.  
Compared to the other approaches we evaluated, which only require the most recent node embeddings as a $|\V|\times d$ matrix for inference, this is a limitation of our model. 
However, this problem can be alleviated by instead storing a $|\V| \times \omega$ matrix of random walks for each node. 
In this way, the edge to test can have its features appended to the relevant walk, and inference can be calculated with a single forward pass.
This requires roughly the same amount of memory; however, it comes with the tradeoff of not accounting for changes in topology over time. 
To be robust, this solution would require a periodic resampling of walks. 
\section{Related Work}
\label{sec:related}

\textbf{Graph analysis for IDS}:
Graphs provide a particular advantage in detecting sophisticated actors in a compromise
because of their ability to encode and navigate heterogeneous relationships in a large
dataset. The connected nature of graph structure allows defenders to detect behavioral
patterns that are anomalous even when they are attempting to disguise themselves
amongst legitimate actions.

Previous work has been done to represent security data as graphs in order to identify
anomalous events. The authors of Argus ~\cite{argus} refer to these as Graph Security
Analytics or GSAs. They can take the form of Host Level provenance graphs, Network
Communication graphs, Authentication graphs, etc. depending on the nature and
granularity of the data available. Once the graph representation has been constructed, they are
then sampled in order to generate tensors for analysis.

Solutions such as Pikachu~\cite{pikachu}, Rabbani et al.~\cite{rabbani2024}, and Anubis~\cite{anubis} generate temporal random walks through the graph in order to generate token sequences informed by the local topology around each node. 
Prior works SIGL~\cite{sigl} and Bowman et al.~\cite{bowman2020} also generate embeddings from random walks through their graphs, though they do not use a temporal constraint.

An alternate approach has been to apply Graph Neural Networks (GNNs) to these different security
graphs in order to learn an embedding representation. 
In Euler~\cite{euler}, the authors train a GNN using snapshots of a network communication graph in order to generate topological embeddings. 
Argus~\cite{argus} builds on this by changing the encoding of the GNN in an attempt to focus on neighborhood characteristics rather than strict graph structure. 
Jbeil ~\cite{jbeil} performs message passing between nodes of a continuous time dynamic graph.
HetGLM ~\cite{hetglm} also uses a GNN as part of the pipeline to generate node embeddings.

\textbf{Graph Foundation Models}
Foundation models have proven very successful in natural language tasks in the form of large language models (LLMs). 
There are now seemingly endless options of LLM families~\cite{llm_survey} that can be fine-tuned or specially trained for countless tasks. 
However, though there are many models under this umbrella, the primary difference between them is how they were trained, and using what data. 
Architecturally, very little has changed about these models since their introduction. 
For our purposes, LLMs can be broadly organized into two classes: GPT-based~\cite{gpt} and BERT-based~\cite{bert}.
GPT-based models are trained to predict the \textit{next} token in a sequence, while BERT-based models are trained to predict a masked token anywhere in a sequence. 

As LLMs revolutionized the NLP space, it was only a matter of time before someone thought to apply this approach to graphs. 
Graph-BERT~\cite{graphbert} is an approach that uses subgraph sampling to pretrain a GFM to reconstruct node features, or local topologies. 
They show that the resulting model can transfer to node and graph classification tasks with or without fine-tuning. 
This work points out the utility of using sequence encoders for GFMs as they are unconstrained by graph size, given informative node features as input tokens. 
GraphGPT~\cite{graphgpt}, not to be confused with~\cite{fakegraphgpt, fakegraphgpt1}, which use the same name, despite its title, also shows that BERT is the most effective training approach for graph foundation models. 
GraphGPT tests various LLM training pipelines that pretrain on sequences of ShaDowKHop~\cite{shadowgnn} subgraph samples to predict next tokens (node and edge features). 
They showed that their approach could then be fine-tuned for link prediction in large knowledge graphs. 
This work is the most informative for ours, though we use the more efficient random walk-based sampling strategy. 
Relphormer~\cite{relphormer} is a sequence encoding architecture specifically designed for knowledge graphs. 
Subgraphs are sampled from a large graph, provided to a transformer as a list of tokens, then the adjacency matrix is included in the transformer step when the attention matrix is generated. 
Though they do not use this approach for a GFM, the architecture they propose could be used in this domain. 
\section{Conclusion}
\label{sec:conclusion}

In this work, we present {\name}: an anomaly-based intrusion detection system designed using the LLM paradigm. 
We extend the traditional random walk-based approaches that use shallow neural networks for node embedding, to modern deep learning architectures and training methods. 
In doing so, we produced a highly precise link prediction system. 
This system, when applied to lateral movement detection in enterprise networks, can detect anomalous authentications, flows, and connections with higher precision than the current state-of-the-art in this field. 
We demonstrated that pretraining {\name} on its own produces highly precise models, and that by finetuning the models, we can improve precision by as much as 580\%. 
Our experiments demonstrated that this approach is robust to missing data, and continues to improve with additional rounds of training. 
This system adapts well to the large amounts of data present in cybersecurity logs, trains quickly, and detects lateral movement with high precision.

\bibliographystyle{IEEEtran}
\bibliography{bib}

\cleardoublepage
\appendix
\section*{Base Model Size Ablation}
In our comparisons to prior works in network anomaly detection, we wanted to keep comparisons fair, and use a similar number of parameters to the preexisting models. 
As such, we only used the ``tiny" variant of BERT, which uses about 2.5M parameters. 
However, we wish to test the scaling laws that have been observed in such models as well. 
Without comparing to the smaller models used in prior works, we present a study on how increasing the size of the FM affects the scores it is able to attain. 

\begin{table}[!htp]
\centering
\scriptsize
\caption{Different sizes of BERT}\label{tab:bert_size}
\begin{tabular}{lrrrrr}\toprule
&Hidden Size &Layers &Attention Heads &Parameters \\\midrule
Tiny &128 &2 &2 &2.5M \\
Mini &256 &4 &4 &7.0M \\
Medium &512 &8 &8 &21M \\
Baseline &768 &12 &12 &98M \\
\bottomrule
\end{tabular}
\end{table}

On the OpTC and UNSW datasets, increasing the size of BERT lead to a sharp decrease in all metrics. This is likely because these datasets have very few unique tokens to learn, and the model quickly overfits. 
The LANL dataset, however, has 15.7k unique tokens, which is more comparable to the vocabulary used by LLMs~\cite{vocab_sizes}. 
We observe that when trained and evaluated on the LANL dataset, increasing the model size led to improvements up to a point. 
The best performing model was BERT-medium, which achieved a 0.90 AP score. 
Adding additional parameters with BERT-baseline resulted in a slightly lower AP, and required a full week to pretrain. 

\begin{table}[!htp]
\centering
\scriptsize
\caption{CyberGFM Metrics and Runtime for Varied BERT Sizes}\label{tab:size_results}
\begin{tabular}{lrrrr}\toprule
    &AUC &AP &Pretraining time \\\midrule
    Tiny &\textbf{0.9994} &0.7600 &6137s \\
    Mini &0.9993 &0.8008 &27270s \\
    Medium &0.9986 &\textbf{0.9005} &102900s \\
    Baseline &\textbf{0.9994} &0.8732 &613000s \\
\bottomrule
\end{tabular}
\end{table}

\begin{figure}[htbp]
    \centering
    \includegraphics[width=\linewidth]{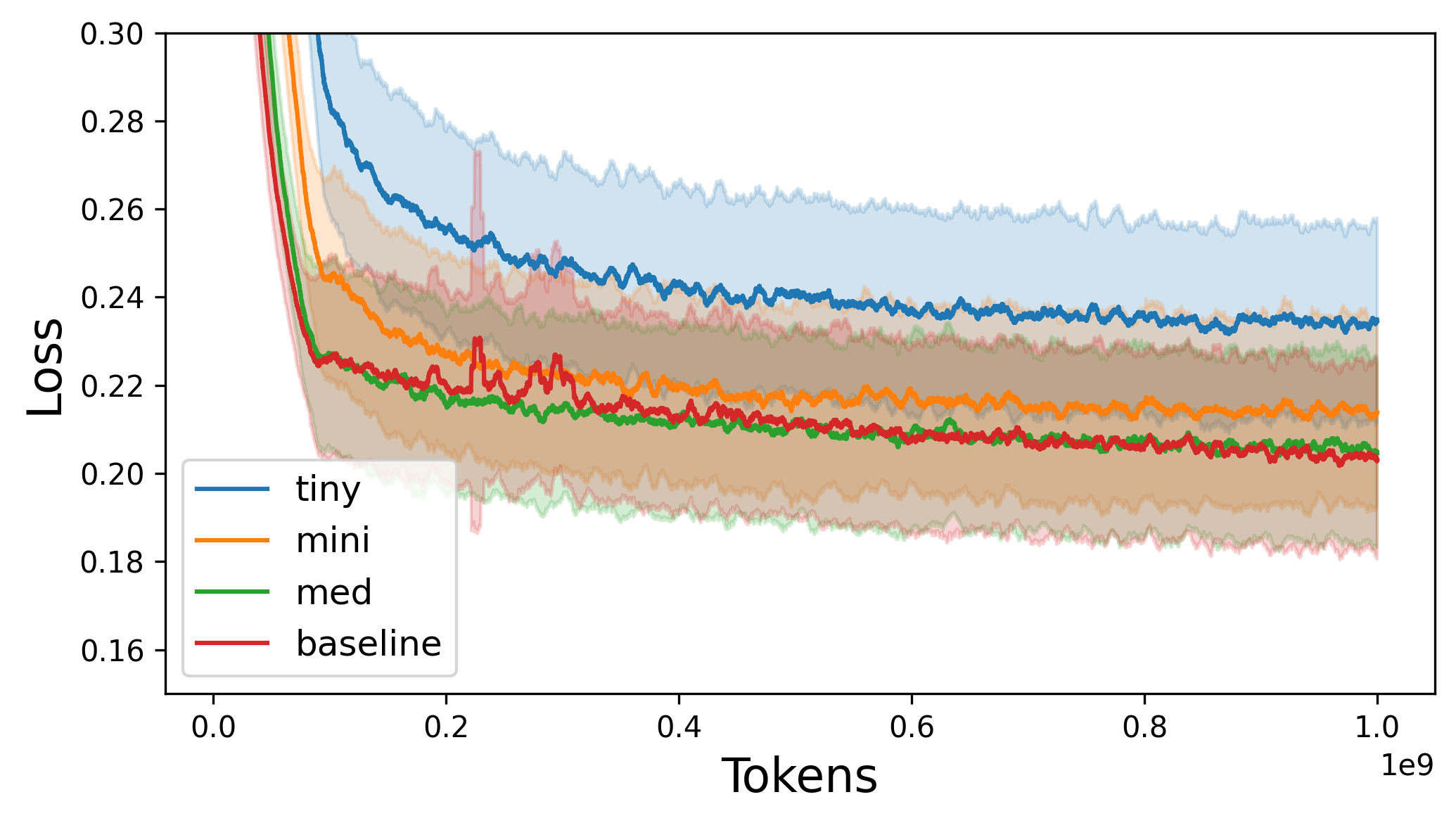}
    \caption{Loss during pretraining for different BERT sizes}
    \label{fig:loss_v_size}
\end{figure}

The scaling laws observed in natural language do not appear to affect graphs as strongly, at least at this scale. 
This is further evidenced by observing the change in loss over the course of model training. 
Figure~\ref{fig:loss_v_size} shows that the difference in loss between the Tiny model and the others is significant. 
However, the difference in loss between the Baseline and Medium models is negligible. 

Recent work~\cite{vocab_to_model_size} has shown that foundation models have an optimal vocabulary size that increases proportionally to the number of parameters up to some bound. 
Because the vocabulary size of our models is bound to the number of nodes in the graph, we cannot freely change this variable. Instead, we change the number of parameters. 
The results of this experiment point to smaller models being more optimal for this particular task, on this specific dataset. 
However, as our approach is applied to larger graphs, larger models will continue providing more benefits. 

\end{document}